\documentclass[lettersize,journal]{IEEEtran}

\usepackage{microtype}                 
\PassOptionsToPackage{warn}{textcomp}  
\usepackage{textcomp}                  
\usepackage{mathptmx}                  
\usepackage{times}                     
\usepackage{cite}                      
\usepackage{tabu}                      
\usepackage{booktabs}                  
\usepackage{calc}
\usepackage{subcaption}
\PassOptionsToPackage{hyphens}{url}
\usepackage{hyperref}

\usepackage{algorithm}
\usepackage{algorithmicx}
\usepackage{algpseudocode}
\usepackage{amsmath}
\usepackage{enumitem}
\usepackage{euscript}
\usepackage{wrapfig}
\usepackage[normalem]{ulem}
\usepackage{graphicx}
\usepackage{flushend}
\usepackage{multirow}
\usepackage{makecell}
\usepackage{xcolor}

\newcommand{\myparagraph}[1]{\mbox{\ } \newline \noindent \textbf{#1}}
\renewcommand{\paragraph}[1]{\vspace{-2.5mm}\myparagraph{#1}}

\newif\ifshowcomments
\showcommentstrue

\ifshowcomments
\newcommand{\TODO}[1]{{\color{red}{[TODO: #1]}}}

\newcommand{\phil}[1]{{\color[rgb]{0.9,0.1,0.1}{#1}}}
\newcommand{\od}[1]{{\color{red}#1}}
\newcommand{\dl}[1]{{\color{magenta}#1}}
\newcommand{\mf}[1]{{\color{blue} [MF: #1]}}
\newcommand{\jdfcomment}[1]{{\small\textcolor{blue}{jdf: #1}}}
\newcommand{\jdf}[1]{{\textcolor{orange}{#1}}}
\newenvironment{jdfenv}{\bgroup\color{orange}}{\egroup}
\newcommand{\cx}[1]{{\textcolor{black}{#1}}}
\newcommand{\cxcomment}[1]{{\textcolor{orange}{[xin: #1]}}}

\else
\newcommand{\TODO}[1]{}
\newcommand{\revised}[1]{}
\newcommand{\phil}[1]{}
\newcommand{\od}[1]{}
\newcommand{\dl}[1]{}
\newcommand{\mf}[1]{}
\newcommand{\jdf}[1]{}
\newcommand{\jdfcomment}[1]{}
\newcommand{\cx}[1]{}
\newcommand{\cxcomment}[1]{}

\fi
\usepackage{amsmath,amssymb,color}
\usepackage{xspace}
\newcommand{\eg}{e.g.,\xspace}
\newcommand{\ie}{i.e.,\xspace}
\newcommand{\ourmethod}{VIDP\xspace}
\algrenewcommand\algorithmicrequire{\textbf{Input:}}
\algrenewcommand\algorithmicensure{\textbf{Output:}}
\begin{document}
\setlength{\abovedisplayskip}{2pt}
\setlength{\belowdisplayskip}{2pt}
\setlength{\abovedisplayshortskip}{0pt}
\setlength{\belowdisplayshortskip}{0pt}

\title{Visualization-Driven Illumination for Density Plots}

\author{Xin Chen, Yunhai Wang, Huaiwei Bao, Kecheng Lu, Jaemin Jo, Chi-Wing Fu, Jean-Daniel Fekete
\thanks{X. Chen and H. Bao are with Shandong University, CN. E-mail:
\{chenxin199634,bhuaiwei\}@gmail.com.}
\thanks{K. Lu is with Renmin University of China, CN. E-mail:  lukecheng0407@gmail.com}
\thanks{J. Jo is with Sungkyunkwan University, KR. E-mail: jmjo@skku.edu.}
\thanks{C.-W. Fu is with the Chinese University of Hong Kong, Hong Kong, CN. Email: cwfu@cse.cuhk.edu.hk}
\thanks{J.-D. Fekete is with University Paris-Saclay, CNRS, Inria, LISN, FR. E-mail: Jean-Daniel.Fekete@inria.fr}
\thanks{Y. Wang is with Renmin University of China, CN, and is the corresponding author. E-mail: wang.yh@ruc.edu.cn}

    
}

\markboth{Submitted to IEEE Transactions on Visualization and Computer Graphics}%
{Shell \MakeLowercase{\textit{et al.}}: A Sample Article Using IEEEtran.cls for IEEE Journals}


\maketitle

\begin{abstract}
We present a novel visualization-driven illumination model for density plots, a new technique to enhance density plots by effectively revealing the detailed structures in high- and medium-density regions and outliers in low-density regions, while avoiding artifacts in the density field's colors.
When visualizing large and dense discrete point samples, scatterplots and dot density maps often suffer from overplotting, and density plots are commonly employed to provide aggregated views while revealing underlying structures. 
Yet, in such density plots, existing illumination models may produce color distortion and hide details in low-density regions, making it challenging to look up density values, compare them, and find outliers.
The key novelty in this work includes (i) a visualization-driven illumination model that inherently supports density-plot-specific analysis tasks and (ii) a new image composition technique to reduce the interference between the image shading and the color-encoded density values. %
To demonstrate the effectiveness of our technique,
we conducted a quantitative study, an empirical evaluation of our technique in a controlled study, and two case studies, exploring 
twelve datasets with up to two million data point samples.  


\end{abstract}

\begin{IEEEkeywords}
Density plot, Illumination, Shading, Image composition
\end{IEEEkeywords}

\begin{figure*}[!t]
  \centering
  \includegraphics[width =0.995\linewidth]{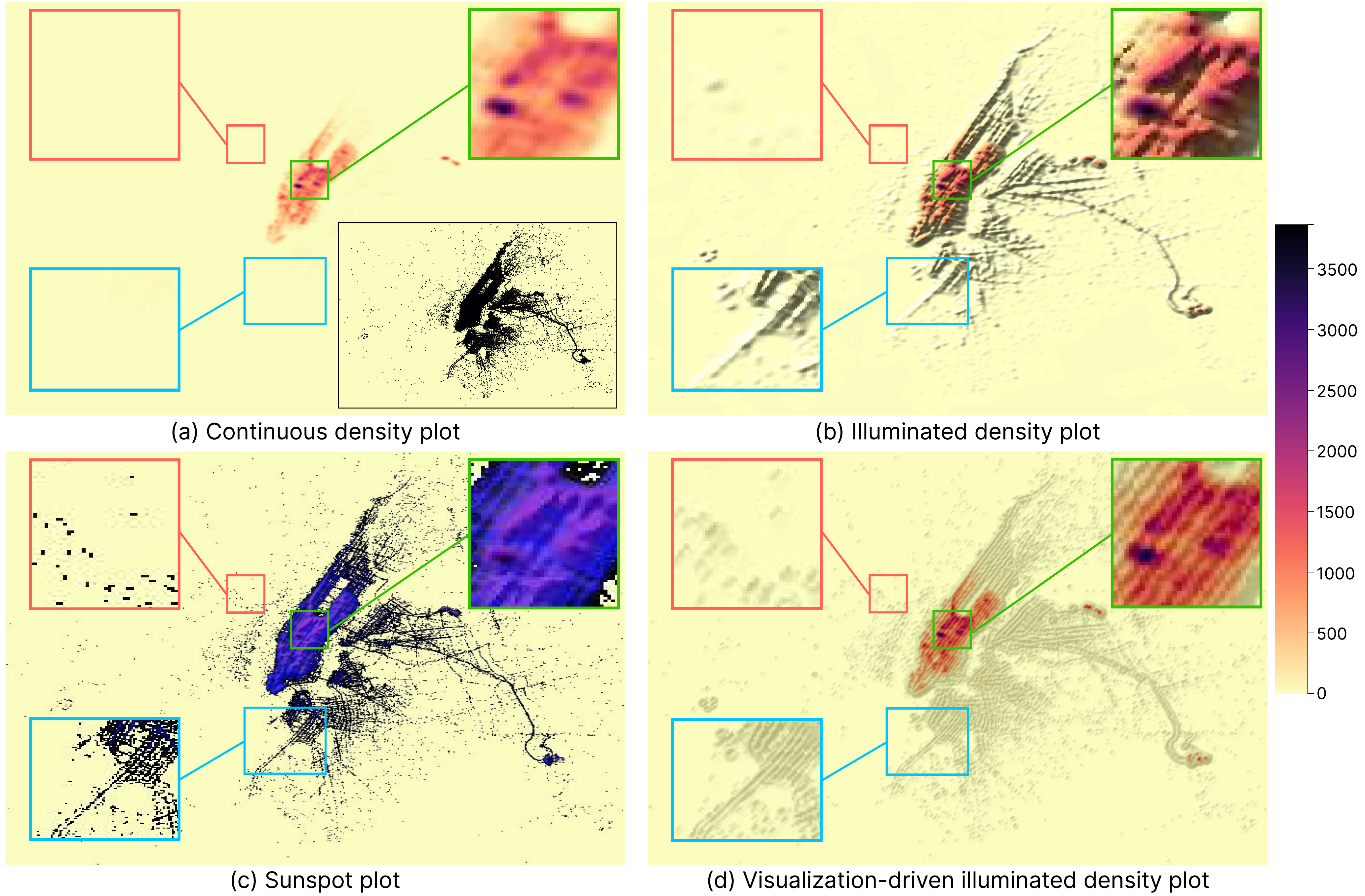}
  \vspace{-6mm}
  \caption{
Different methods for visualizing two million-point samples of the ``New York TLC Trip'' dataset (bottom right in (a) shows the raw plot).
Our technique is shown in (d).
The \cx{green} box represents a high-density (HD) region,
the red box shows a low-density (LD) one,~\ie outliers, 
and the \cx{blue box} has medium-density (MD).
The HD variations are faithfully revealed in (a,d), but (b,c) hinder the perception of absolute density values by using colors absent from the colormap (brown and blue).
On the other hand, LD outliers are revealed in (c,d) and hidden in (a,b).
Finally, the MD structures are revealed in (b,c,d) and hidden in (a).
Our Visualization-driven Illuminated Density Plot (\ourmethod) not only shows the density variations using colors similar to (a) but also reveals MD structures and LD outliers.
}
    \vspace{-5mm}
    \begin{subcaptiongroup}
            \phantomcaption\label{fig:teaser:a}
            \phantomcaption\label{fig:teaser:b}
            \phantomcaption\label{fig:teaser:c}
            \phantomcaption\label{fig:teaser:d}
            \phantomcaption\label{fig:teaser:e}
            \phantomcaption\label{fig:teaser:f}
    \end{subcaptiongroup}
\label{fig:teaser}
\end{figure*}

\section{Introduction}\label{sec:intro}


Scatterplots are among the most effective techniques for visualizing discrete data points in 2D. 
Yet, for large and dense data, scatterplots suffer from overplotting, where visual clutter obscures the data distribution (\autoref{fig:teaser:a}).
A few alternatives have been proposed to overcome such a limitation;~\eg, instead of visualizing each point as a single mark, \emph{2D density plots} (also called \emph{density maps} and \emph{heatmaps}, we use the term {\em density plots\/}) color-encode the aggregated density of the data points, usually smoothed using 
Kernel Density Estimation (KDE).


Density plots could effectively reveal global patterns (\eg trends and clusters), especially in high-density regions. However, they often neglect less dense, local patterns such as outliers in low-density regions~\cite{sarikaya2018scatterplots} (the red box in \autoref{fig:teaser:a}). They also sometimes hide important visual structures (hereafter ``structures'') in medium-density regions, \ie meaningful local density variations~\cite{gibson1950perception} (the \cx{blue box} in \autoref{fig:teaser:a} \cx{vs. \ref{fig:teaser:b},\ref{fig:teaser:c},\ref{fig:teaser:d}}). This is because of the limited intensity resolution of existing displays and our vision system: assigning indistinguishable colors to the medium- and low-densities hinders several important analysis tasks such as \emph{identifying anomalies}~\cite{sarikaya2018scatterplots}.

Several density plot alternatives have been proposed to enhance the visibility of global and local patterns simultaneously. 
By representing a density field as an illuminated height field using Phong shading~\cite{willems2009visualization}, 
the Illuminated Density Plot (IDP) provides a clearer depiction of the main structures in medium- and high-density regions (\autoref{fig:teaser:b}).
However, the shading operation may introduce artifacts, such as altering beige and dark brown in~\autoref{fig:teaser:a}, 
making it challenging to look up and compare density values. Additionally, some outliers in low-density regions, particularly those in the red box of \autoref{fig:teaser:b}, remain hidden.
Explicitly overlaying outliers on the IDPs,  Trautner et al.~\cite{trautner2020sunspot} introduced Sunspot Plots (SUPs), which smoothly blend discrete data points with IDPs by using two types of KDE kernels.
However, 
SUPs may introduce colors falling outside of the color map, such as the blue color in \autoref{fig:teaser:c}, which hampers density value look-up and interpretation.
This issue can be attributed to the interference between colored patterns and shaded structures~\cite{ware2010visual}[pp. 83].


The limitations of the previous techniques motivated us to revisit the illuminated density plots, in which the additional illumination should inherently support density-plot-specific analysis tasks.
Based on a survey on scatterplot tasks~\cite{sarikaya2018scatterplots} and user study results~\cite{trautner2020sunspot}, we derived three design requirements that a good density plot design should meet:
\begin{description}[leftmargin=!,labelwidth=\widthof{\bfseries DR1:},nosep]
\item[\textbf{DR1:}]  revealing the detailed structures in high- and medium-density regions;
\item[\textbf{DR2:}]  maintaining the visibility of outliers in low-density regions; and
\item[\textbf{DR3:}] producing less color distortion to support accurate lookup and comparison of absolute density values as much as possible.
\end{description}
In this article, we present a novel density plot design, Visualization-driven Illuminated Density Plots (\ourmethod), aiming to fulfill all three requirements. Here, ``visualization-driven'' means that the illumination model is customized to be perceptually effective in visual analysis tasks such as value lookup, comparison, and outlier identification, rather than directly using the ones developed by the computer graphics community.
To the best of our knowledge, this is the first technique that considers the three requirements simultaneously.



Instead of applying the well-known Phong shading model~\cite{Phong1975illumination} to density field rendering, 
we explore a structure-enhancing shading model 
that adheres to some principles of manual relief shading~\cite{rusinkiewicz2006exaggerated}, such as shading along ridges and valleys, omitting shadow and specular reflections, and maximizing the overall contrast. To combine the shading image from this model with the colored density image, we further propose a new composition scheme 
that adjusts only the luminance of the density field's colors to minimize the
interference between the shaded image and color-encoded density value.
As shown in \autoref{fig:teaser:d}, the resulting density plot with valid colors reveals the main structures and major outliers. \autoref{table:compare-method} compares our density plot with prior techniques 
on three levels of densities.

We evaluate our approach by comparing it with state-of-the-art density visualizations (\eg SUP~\cite{trautner2020sunspot}) using ten datasets with up to two million data point samples.
First, we conduct a quantitative evaluation with an established color distance measure~\cite{sharma2005ciede2000}, showing that \ourmethod significantly reduces 
color distortions compared with others.
Second, we evaluate our approach in terms of density-driven analysis tasks\footnote{Experimental data and analysis code are included with the submission as supplemental materials and are available at \url{https://osf.io/5xpsw/?view_only=0445046dad574d4a90d7138e94547ada}.} in a controlled user study. Our \ourmethod achieves comparable results in preserving densities and outliers, while better revealing local patterns.
In addition, we conduct two case studies to demonstrate the effectiveness of \ourmethod, showing that interactively adjusting the parameters in our method enables viewers to explore and highlight structures of interest.
To summarize, our main contributions are as follows:
\begin{itemize}[nosep]
  \item We propose a visualization-driven illumination model that inherently supports density-plot-specific analysis tasks.
  \item We propose a new image composition technique to reduce the interference between the shaded image and color-encoded density values.
  \item We empirically 
  evaluate our techniques in a controlled user study on ten datasets and present two case studies to demonstrate their effectiveness. 
\end{itemize}

\begin{table}[!t]
	\vspace*{-1mm}
	\centering \resizebox{0.95\linewidth}{!}{\begin{tabular}{@{}l|c@{\hspace{3mm}}c@{\hspace{3mm}}c@{}}
			\toprule
			 &  high-density & medium-density & low-density \\
			\midrule
			{CDP} & no color distortion & barely visible & invisible  \\
			{IDP~\cite{willems2009visualization}} & some color distortion & visible & invisible  \\
			{SUP~\cite{trautner2020sunspot}}  &  strong color distortion & visible & visible   \\
			{\ourmethod} & weak color distortion & visible & visible  \\
			\bottomrule
			\end{tabular}}
		\vspace*{-1.5mm}
		\caption{
			Comparing our visualization-driven illuminated density plots (VIDP) with prior techniques on capabilities to visualize 
   regions of 
      different levels of densities, as demonstrated in \autoref{fig:teaser}.
		}
		\vspace*{1.5mm}
		\label{table:compare-method}
\end{table}

\section{Related Work}

\subsection{Density Visualization}
Density plots are one of the most common ways to combat the overplotting problem in scatterplots.  Below, we discuss the generation, enhancement, and evaluation of density plots.

\vspace*{-1mm}
\paragraph{Density Plots Generation}.
Opacity adjustment and density estimation are two common approaches to generate density plots.
Assigning suitable opacity values to data points can produce semi-transparent visualizations that better reveal high-density regions.
Matejka et al.~\cite{matejka2015dynamic} present a user-driven model for setting opacity values based on data distribution and crowd-sourced responses.
Micallef et al.~\cite{micallef2017towards} suggest optimizing the opacity, mark size, and other visual properties together in line with the given data and the analysis task. Yet, alpha blending is limited to a few layers and it is hard to distinguish between varying densities in the generated density plots.

Instead, density estimation explicitly computes a smooth density field, which is often shown as a color-coded plot. 
Usually, the densities are estimated by kernel density estimation (KDE)~\cite{silverman1986density}, which sums the contribution of discrete data samples around each screen pixel based on a kernel function:
\begin{align}
    KDE_h(x) & = \frac{1}{nh} \sum_{i=1}^n K(\frac{x-x_i}{h}),
\label{eq:kde}
\end{align}
where $X = \{x_1, x_2, ..., x_n | x_i \in \mathbb{R}^2\}$ is the bivariate dataset, $K$ is a kernel function, and $h$ is a \emph{bandwidth} parameter.
For convenience, we refer to the resulting plots as continuous density plots (CDP).
Popular kernel functions are the Gaussian and Epanechnikov kernels, which perform well in showing global data distribution in high-density regions.
Heer~\cite{heer2021fast} found that combining linear binning and Deriche's approximation efficiently produces pixel-perfect Gaussian KDE.
For continuous scientific data, Bachthaler and Weiskopf~\cite{bachthaler2008continuous} compute the continuous density field by mapping the spatially continuous input data to the range domain with interpolation between associated data values.
However, the smoothed density field might reduce the local variability, so some important structures could be missed (\autoref{fig:teaser:a}).
%
In this work, we extend such a technique to better account for the major prominent structures and local density variations and also to allow for interactively exploring
detailed vs.\ smoothed results.


\begin{figure*}[!t]
	\centering
	\includegraphics[width=0.99\linewidth]{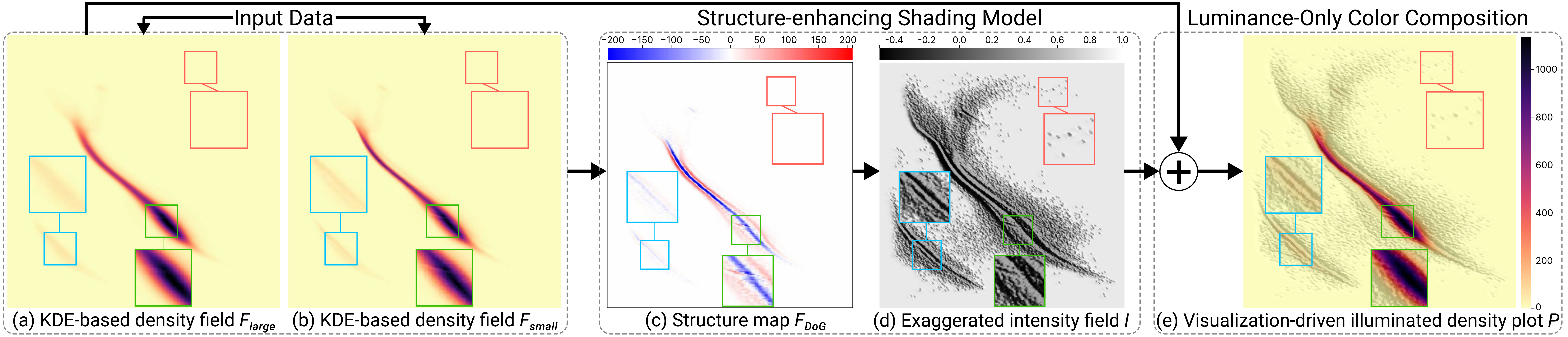}
	\caption{
		The overall pipeline of our visualization-driven illuminated density plots:
        Given a set of discrete points, we first compute the KDE-based density fields (a) $F_{\textit{large}}$ with a large Gaussian kernel and (b) $F_{\textit{small}}$ with a small kernel. 
        Then, we (c) take their difference to derive the structure map $F_{\textit{DoG}}$ and (d) employ diffuse shading with height exaggeration with an automatic light direction set up to obtain the intensity field $I$.
        By further combining $F_{\textit{large}}$ with $I$ using our luminance-only color composition, we can generate (e) the Visualization-driven Illuminated Density Plot (\ourmethod).
	}
	\vspace{-6mm}
        \begin{subcaptiongroup}
            \phantomcaption\label{fig:enhancement_pipeline:a}
            \phantomcaption\label{fig:enhancement_pipeline:b}
            \phantomcaption\label{fig:enhancement_pipeline:c}
            \phantomcaption\label{fig:enhancement_pipeline:d}
            \phantomcaption\label{fig:enhancement_pipeline:e}
            \phantomcaption\label{fig:enhancement_pipeline:f}
        \end{subcaptiongroup}

	\label{fig:enhancement_pipeline}
\end{figure*}

\paragraph{Density Plots Enhancement}.
Traditional continuous density plots cannot effectively depict both local variations and outliers. 
To preserve outliers, 
Splatterplots (SPP)~\cite{mayorga2013splatterplots} explicitly show sub-sampled data points in sparse regions and use closed smooth contours in dense regions. As SPP are developed for showing clusters in multi-class scatterplots, they assign the same color to high-density regions enclosed by contours and discrete data points of the same class in sparse regions, hiding the density variations and leading to ambiguity~\cite{kindlmann2014algebraic}: identical visual intensities might correspond to different effects. 

On the other hand, Willems et al.~\cite{willems2009visualization} used the photorealistic Phong shading~\cite{Phong1975illumination} to a height map, combining two KDE density fields with different bandwidths to reveal local structures on the colored density plot. However, such an illuminated density plot (IDP) cannot clearly show outliers (red box in \autoref{fig:teaser:b}), 
so the authors adopted a specific approach for trajectory data to make the outliers pop up, which does not apply to general bivariate datasets.
Furthermore, IDP distorts the original colors of the density plot (\autoref{fig:teaser:b}) and prevents users from looking up and comparing density values based on the color map.

SUP~\cite{trautner2020sunspot} computes two density fields by performing KDE on the data with 
two different types of kernels.
After rendering these density fields using different colormaps, they combine the two 
rendered density fields adaptively together and exploit the shading and shape cues to maintain the relative density differences. The
recently-proposed honeycomb plots~\cite{10.2312:vmv.20221205} further improve the shading.
In doing so, most outliers can be more clearly shown.
Yet, their evaluation shows that both techniques cannot improve the accuracy of estimating density values. Also, they still cannot effectively convey local density variations (\cx{the green} box in \autoref{fig:teaser:c} \cx{vs. \ref{fig:teaser:b},\ref{fig:teaser:d}}) and also suffer from the ambiguity issue as Splatterplot (black color in 
the red and \cx{green} boxes in \autoref{fig:teaser:c}).
Besides, SUP's color blending may lead to significant color distortions (\eg the blue and purple colors in the \cx{green} box in \autoref{fig:teaser:c}).


\paragraph{Density Plots Evaluation}.
Sarikaya and Gleicher~\cite{sarikaya2018scatterplots} list a set of scatterplot-specific tasks and discuss the effectiveness of scatterplots, contour plots, and SPP in support of the aggregation-level tasks. Recently, Trautner et al.~\cite{trautner2020sunspot} conducted a user study to compare five visual designs with two tasks (density estimation and comparison) and found that SUP performs similarly to the KDE-based continuous density plots. In this work, we extend this study with four variants of visual design and three tasks.

\subsection{Illumination Models}
Various illumination models have been developed~\cite{hughes2014computer}. Here, we restrict our discussion to non-photorealistic models~\cite{gooch2001non} for depicting shape details and important structures rather than creating realistic images.
For example, Cignoni et al.~\cite{cignoni2005simple}  suggest performing diffuse shading with a sharpened normal field to increase the contrast at corners. 
Rusinkiewicz et al.~\cite{rusinkiewicz2006exaggerated} proposed a multi-scale shading model to convey both overall shape and fine-scale detail by computing the normal field at different scales and varying the light source to maximize the contrast. 
Inspired by these techniques, we propose a visualization-driven illumination model for density plots rather than simply applying the one developed by the computer graphics community. 
Specifically, we compute a normal field to highlight ridges and valleys in the density field and apply diffuse shading with 
an optimized
lighting direction to maximize the contrast within such structures. 
To ensure accuracy in density value estimation, we further consider minimizing the artificial colors resulting from the composition of the shading image and the input colored density field. We show that the resulting density plots better support specific analysis tasks such as value estimation and outlier identification via a controlled study and two case studies.



\section{Visualization-driven Illuminated Density Plots}
Our goal is to enhance density plot visualizations, revealing the underlying structures and outliers as much as possible while ensuring the perception of color-encoded density values.
We approach this problem by first designing a 
structure-enhancing shading model as illustrated in \autoref{fig:enhancement_pipeline}. 
Here, we create a shading intensity field to capture prominent structures in the density plot (\autoref{subsect:shadingmodel}), then develop a color composition method to further integrate the shading intensity field into the luminance channel of the colored density plot to reduce color distortions (\autoref{subsect:colorcomposition}).
Last, we conduct a parameter analysis to explore how different parameters affect the quality of illumination (\autoref{subsect:parameteranalysis}).
Note that for simplicity, we only use one example dataset in this section; more results can be found in the supplementary material.

\subsection{Structure-enhancing Shading Model}\label{subsect:shadingmodel}
Kernel density estimation is a widely-used approach for characterizing the high-density regions~\cite{unwin2006graphics,bachthaler2008continuous}, but local details are often poorly retained in the results~\cite{trautner2020sunspot}.
In contrast, using a shading field~\cite{van2001enridged} can easily discern structures, so we are motivated to develop a shading model to depict structural information.
To meet \textbf{DR1} and \textbf{DR2}, we first compute a structure map to capture important structures,~\ie
outliers and detailed structures.
While there are no clear definitions of outliers, we follow Mayorga et al.~\cite{mayorga2013splatterplots} in treating them as data points in low-density regions, but we do not set a specific density threshold.
Meanwhile, we regard high-frequency spatial variations as structures, because they can identify informative local patterns at a fine scale.


Inspired by the principles of manual relief shading~\cite{rusinkiewicz2006exaggerated}, i.e., shading along ridges and valleys, omitting specular reflections, and maximizing overall contrast, we create a shading intensity field containing structures at the detail level in three steps.
First, we construct a structure map to capture important structures such as the ridges and valleys 
(\autoref{fig:enhancement_pipeline:c}).
Then, we adopt diffuse shading with height exaggeration to render structures in the structure map, allowing for emphasizing local details in regions of interest  (\autoref{fig:enhancement_pipeline:d}).
Finally, we take a data-driven approach to automatically set up the lighting 
to provide maximal overall contrast.

\begin{figure}[!t]
	\centering
	\includegraphics[width=0.94\linewidth]{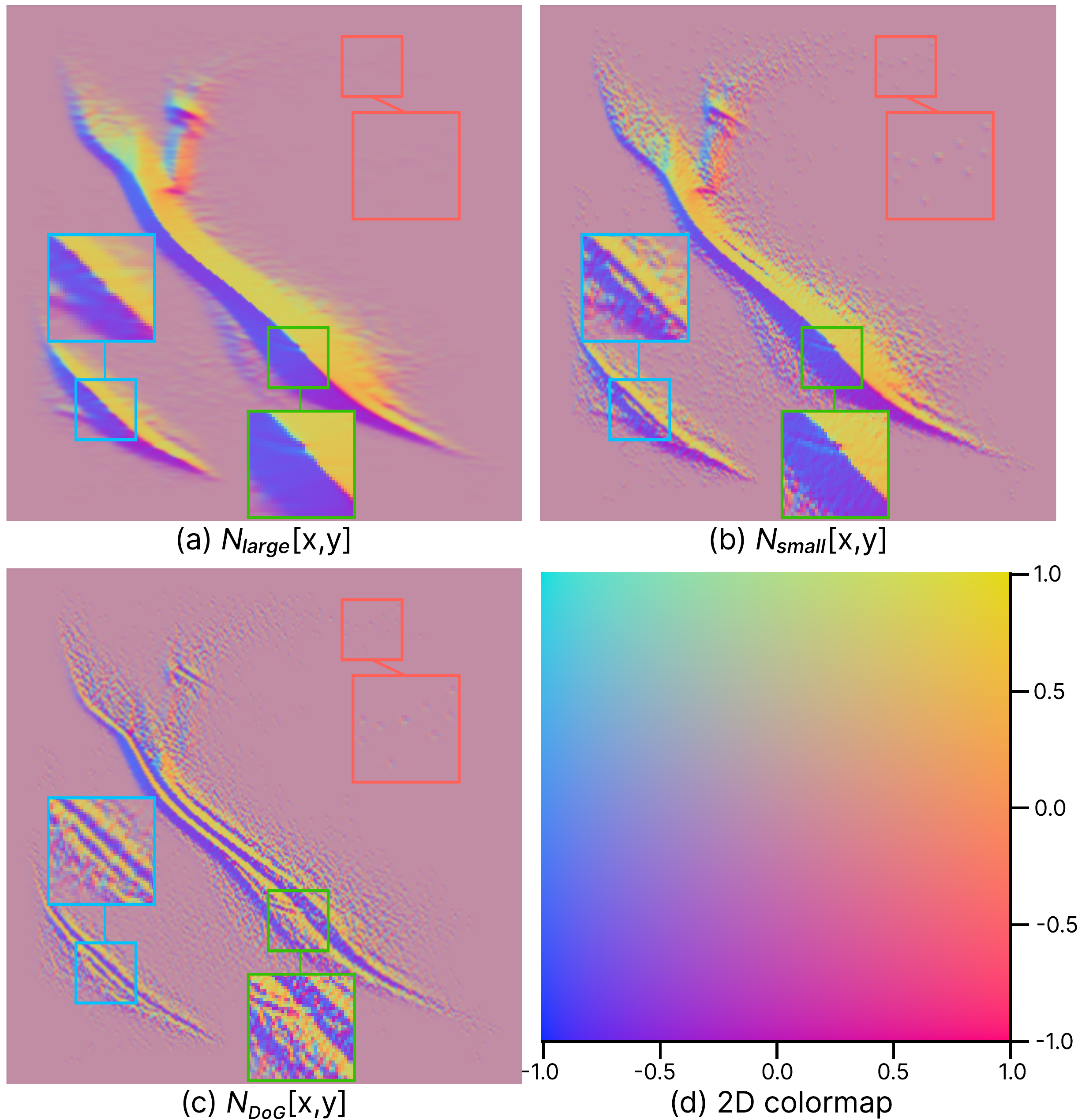}
	\vspace{-1mm}
	\caption{
        Comparison of (a,b) the normal of the density fields $N_{\textit{large}}$ and $N_{\textit{small}}$ and (c) the structure map $N_{\textit{DoG}}$ for revealing local features, where the $x$- and $y$-components of the normal vectors are visualized using (d) a bivariate color map~\cite{steiger2014visual}. The local structures in the blue and green boxes are clearer in (c).
        }
        \begin{subcaptiongroup}
            \phantomcaption\label{fig:DoG_analysis:a}
            \phantomcaption\label{fig:DoG_analysis:b}
            \phantomcaption\label{fig:DoG_analysis:c}
            \phantomcaption\label{fig:DoG_analysis:d}
        \end{subcaptiongroup}
	\label{fig:DoG_analysis}
\end{figure}

\paragraph{Structure map construction.}
One canonical way to construct the structure map is to analyze an image at different scales created by Gaussian smoothing, usually by measuring the difference in the densities estimated by Gaussians with different kernel sizes~\cite{ip2011saliency},~\ie Difference of Gaussians (DoGs), as follows:
\begin{align}
F_{\textit{DoG}}(x) & = \text{KDE}_{h_{\textit{large}}}(x) - \text{KDE}_{h_{\textit{small}}}(x),
\label{eq:DoG_construction}
\end{align}
%
where KDE$_h$() is the Gaussian KDE (\autoref{eq:kde}) and $h_{large}$ and $h_{small}$ are the large and small bandwidths that jointly determine the scales of structures being extracted. In our implementation, we set them using empirical rules; $h_{\textit{large}}$ is set according to \emph{Silverman’s rule of thumb}~\cite{silverman1986density} and $h_{\textit{small}}$ is set to the reciprocal of the width of the grid evaluated by KDE, because the radius of influence of each point is exactly one grid cell in this situation, making outliers stand out from the background.

After obtaining the structure map $F_{\textit{DoG}}$, we can easily derive its surface normal from the gradient as follows:
\begin{align}
N_{\textit{DoG}} & = \frac{1}{\sqrt{{\partial_{x}F_{\textit{DoG}}}^2+{\partial_{y}F_{\textit{DoG}}}^2+1}} [-\partial_{x}F_{\textit{DoG}}, -\partial_{y}F_{\textit{DoG}}, 1],
\label{eq:compute_normal}
\end{align}
while $N_{\textit{large}}$ and $N_{\textit{small}}$ can be derived similarly.
Compared to the density fields $F_{\textit{large}}$ or $F_{\textit{small}}$, the structure map $F_{\textit{DoG}}$ can better enhance local structures because locally high and low densities become noticeable by the differentiation between different scales of the density field, regardless of the absolute densities.
As a result, $N_{\textit{DoG}}$ helps to create shading along ridges and valleys, following the principles of manual relief shading.
\autoref{fig:DoG_analysis} shows the corresponding $x$- and $y$-components of the normal vectors of \autoref{fig:enhancement_pipeline:a},\ref{fig:enhancement_pipeline:b},\ref{fig:enhancement_pipeline:c}, where the green, blue, and red boxes mark high-, medium-, and low-density regions, respectively. Due to the large absolute densities, $N_{\textit{large}}$ and $N_{\textit{small}}$ are visibly separated into two contiguous parts (yellow and blue), hiding the local variations falling into one part. In contrast, $N_{\textit{DoG}}$ does not look contiguous, embodying local structures such as the two bands from top left to bottom right in the blue box.
In conclusion, the structure map is more effective than the Gaussian density field alone for revealing the detailed structures in high- and medium-density regions (\textbf{DR1}).

\paragraph{Diffuse shading with height exaggeration.}
Ambient light is uniform and lacks directionality, whereas specular light creates concentrated highlights that could distract the visualizations.  They are not chosen
to depict the structural information of $F_{\textit{DoG}}$.
Hence, we exploit Lambertian shading~\cite{lambert1760photometria} 
to produce a visual cue for perceiving the structural details~\cite{horn1981hill}.
Typically, this model assumes an ideal surface that reflects lights uniformly toward all directions in the upper hemisphere of the surface and 
the reflected intensity $I$ obeys the Lambert's cosine law:
\begin{align}
    I = N \cdot L = |N||L| cos\theta = cos\theta,
\label{eq:Lambert_cosine_law}
\end{align}
where $N$ is the unit normal vector,
$L$ is the unit light direction vector pointing from the surface to the light source, and $\theta$ is the angle between vectors $N$ and $L$.
The intensity $I$ decreases as $\theta$ increases and will be negative when 
$\theta>90^{\circ}$,~\eg the shadowed side of a ridge.

\begin{figure}[!t]
	\centering
	\includegraphics[width=0.99\linewidth]{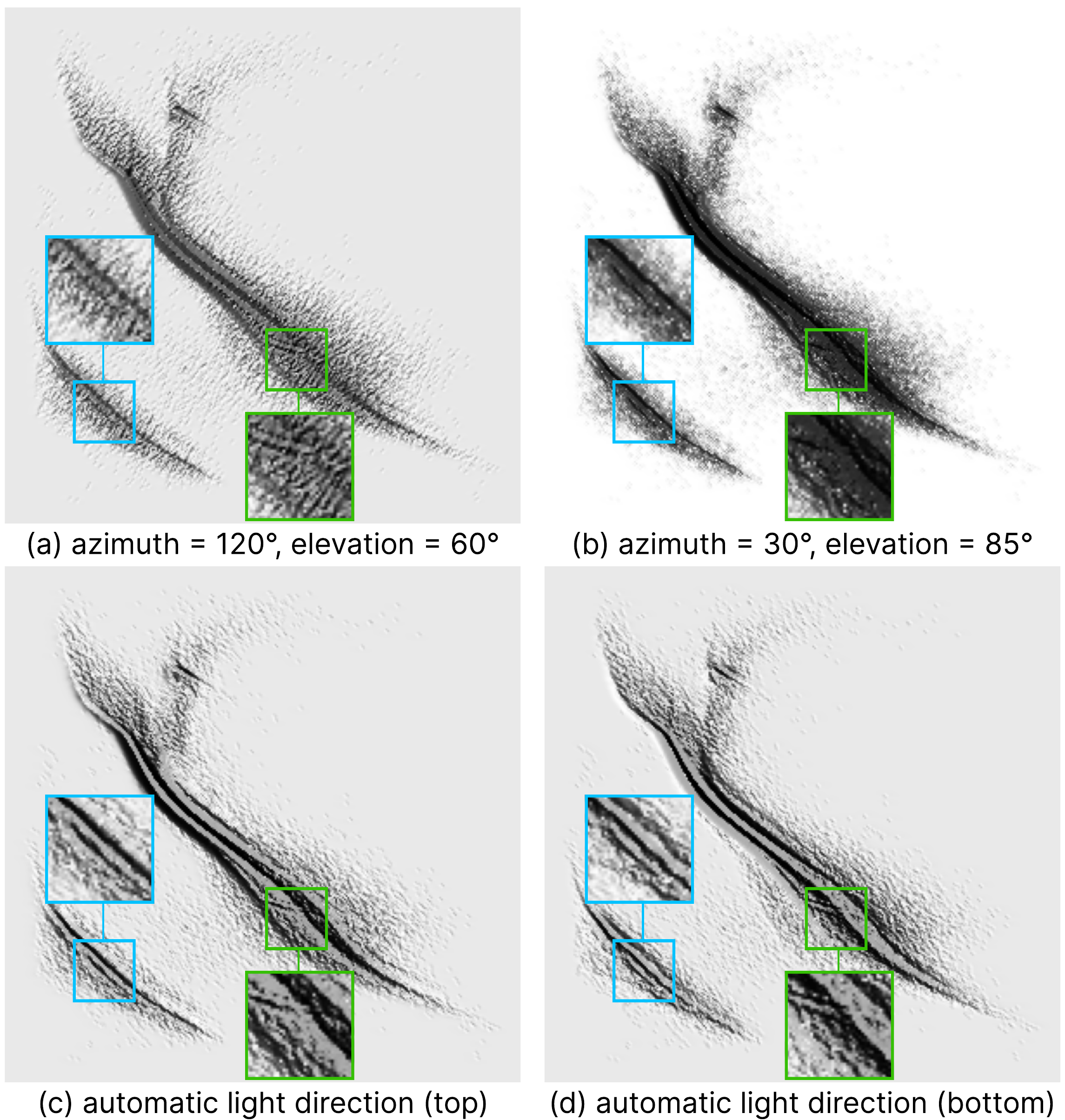}
	\vspace{-5mm}
	\caption{
        Comparison of light directions. (a,b) are unsuitable, while (c,d) are effective. 
        Our automatic illumination setup can find the two appropriate light directions, (c) negative and (d) positive, along the principal vector with the highest variance without a tedious trial-and-error process. 
        Following O'Shea et al.~\cite{o2008assumed}, we choose the one from the top of the plot ($y<0$)
        , which leads to the most accurate estimations of shapes and highlights the structures shown in \autoref{fig:DoG_analysis:c}.
        }
        \begin{subcaptiongroup}
            \phantomcaption\label{fig:light_setup:a}
            \phantomcaption\label{fig:light_setup:b}
            \phantomcaption\label{fig:light_setup:c}
            \phantomcaption\label{fig:light_setup:d}
        \end{subcaptiongroup}

	\label{fig:light_setup}
\end{figure}

Although \textbf{DR1} can be satisfied by just applying diffuse shading to the normal of the structure map $N_{\textit{DoG}}$ as shown in \autoref{fig:light_setup:c}, 
the visibility of low-density outliers remains an issue because the slopes caused by the outliers are often too small. For example, it is hard to perceive outliers in the red box of~\autoref{fig:DoG_analysis:c}.
To address this issue, the height exaggeration method in relief shading~\cite{rusinkiewicz2006exaggerated} suggests multiplying the values in the height field ($F_{\textit{DoG}}$ in our case) by a user-specified factor $\eta$; 
in general, a large $\eta$ improves the visibility of structures in low-density regions.

Putting $\eta$
into \autoref{eq:compute_normal}, 
we obtain
\begin{align}
N_{\textit{DoG}} & = \frac{[-\eta \partial_{x}F_{\textit{DoG}}, -\eta \partial_{y}F_{\textit{DoG}}, 1]}{\sqrt{(\eta \partial_{x}F_{\textit{DoG}})^2+(\eta \partial_{y}F_{\textit{DoG}})^2+1}}.
\label{eq:exaggerated_normal}
\end{align}
This means the $z$-component of $N_{\textit{DoG}}$ is close to being inversely proportional to $\eta$, while the horizontal directions of $N_{\textit{DoG}}$ are
unaffected
by $\eta$. In addition, when the magnitude of the gradient is large, $\eta$ has less impact on the normal,
since the $z$-component is already small.

\begin{figure}[!t]
	\centering
	\includegraphics[width=0.99\linewidth]{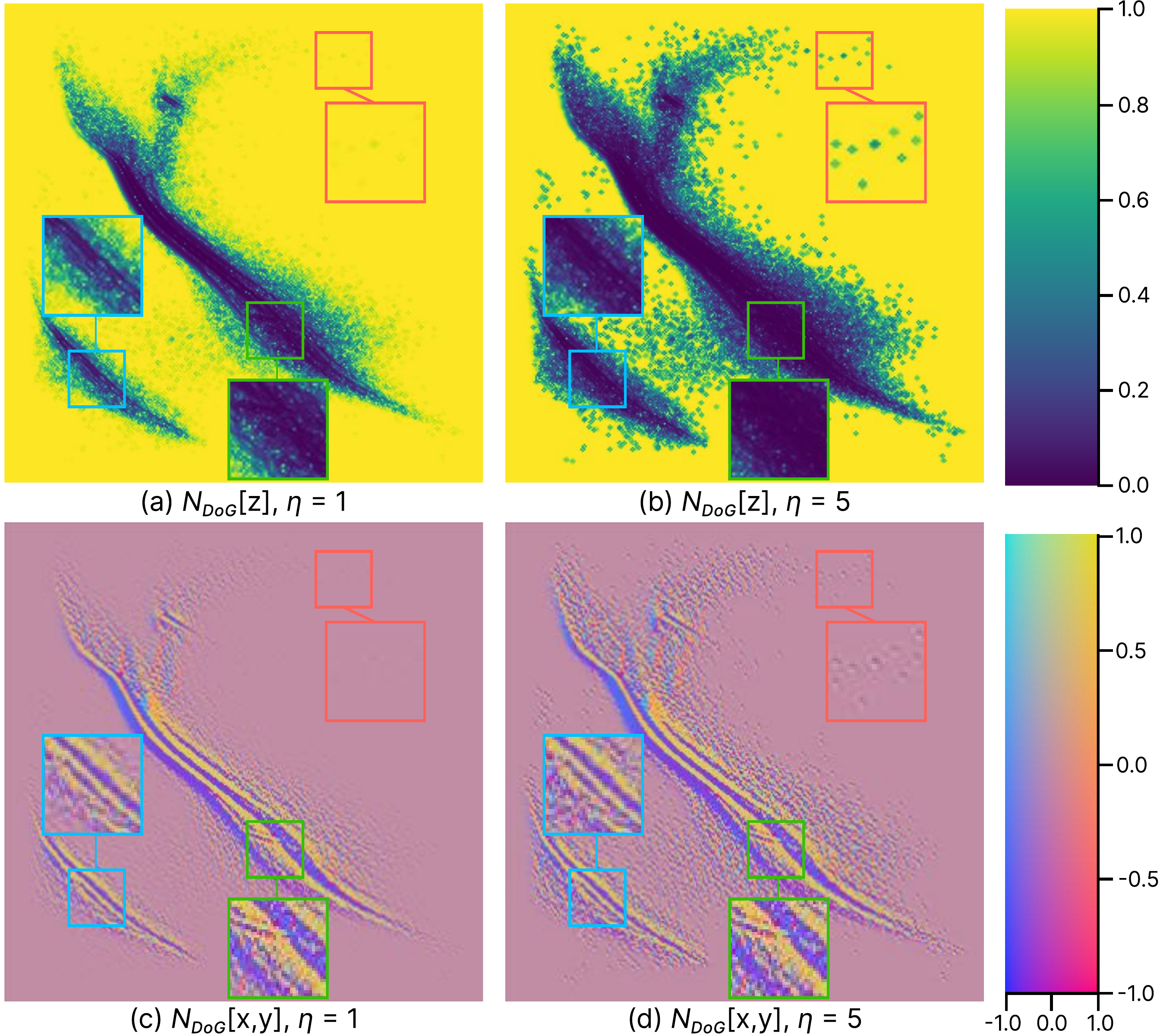}
	\vspace{-5mm}
	\caption{
        The effect of different exaggeration factors (a,c) $\eta=1$ and (b,d) $\eta=5$ on the exaggerated normal $N_{\textit{DoG}}$. (a,b) The $z$-components are visualized by the 1D colormap Viridis~\cite{hunter2007matplotlib} (right), and (c,d) the $x$- and $y$-components are visualized by the same 2D colormap used in \autoref{fig:DoG_analysis} (right). Setting $\eta$ to 5 reveals the outliers in the red box 
        without noticeably changing the
        major structures in the blue and green boxes.
        }
        \begin{subcaptiongroup}
            \phantomcaption\label{fig:eta_comparison:a}
            \phantomcaption\label{fig:eta_comparison:b}
            \phantomcaption\label{fig:eta_comparison:c}
            \phantomcaption\label{fig:eta_comparison:d}
        \end{subcaptiongroup}

	\label{fig:eta_comparison}
\end{figure}

\autoref{fig:eta_comparison} shows an example of the effect of $\eta$ on the different components of $N_{\textit{DoG}}$.
By increasing $\eta$ from 1 to 5, the $z$-components of unit normal vectors are shrunk while the $x$- and $y$-components are expanded correspondingly.
Hence, the outliers in the red box are revealed by smaller $z$-components and larger horizontal magnitudes with $\eta=5$ (\autoref{fig:eta_comparison:b},\ref{fig:eta_comparison:d}), satisfying \textbf{DR2}.
However, increasing $\eta$ is not helpful in the high- and medium-density regions such as the blue and green boxes, where the main structures are shown clearly with $\eta=1$ (\autoref{fig:eta_comparison:c}). Also, a large $\eta$ that amplifies minor variations in these regions may interfere with the perception of major variations, so decreasing $\eta$ can be helpful in some cases.
Hence, we provide an interactive system for users to interactively emphasize local details in regions of interest by brushing and adjusting $\eta$.
An example is given in~\autoref{sec:Case}.

\paragraph{Automatic lighting setup.} \label{sec:auto_light_setup}
Applying an appropriate lighting direction $L$ plays an important role in making the resulting shading ($N_{\textit{DoG}} \cdot L$) more effective in revealing structural information.
For example, \autoref{fig:light_setup:a} shows a result with an unsuitable lighting direction where structures (blue and green boxes) are not effectively shown.
However, manually setting the light direction can be too tedious, so we adapt an automatic 2D lighting setup strategy from volume rendering~\cite{zhang2013lighting} to maximize the overall contrast.
Instead of modeling a quality metric and iteratively refining the light parameters, we collect the statistical information from the image and set up the light using an empirical lighting design model, which is fast, stable, and close to a human design~\cite{zhang2013lighting}.

By computing the mean and spatial variation of normal vectors $N_{\textit{DoG}}$, we obtain statistical information about the structures in $F_{\textit{DoG}}$. We choose to filter out the normal vectors of empty regions of the scatterplot (the normal vectors [0,0,1]), since they do not contain valuable structures; we refer to the remaining normal vectors as $N'$. 
To maximize the shading contrast, we perform Principal Component Analysis (PCA) on the $x$- and $y$-dimensions of $N'$  
to acquire the principal vectors $\left\{v_1, v_2\right\}$ and corresponding variances $\left\{\lambda_1, \lambda_2\right\}$ that
encode the principal directions of the normal vectors.
Note that we do not use the $z$-dimension of the vectors to avoid selecting a high-elevation direction (elevation $\approx90^{\circ}$) because typically, a flat object with small surface details needs a light at a grazing angle to best reveal its shape~\cite{wambecke2016automatic}.
A bad example is shown in~\autoref{fig:light_setup:b}, in which empty regions are assigned the highest luminance value while the main structures are obscured due to indistinguishable low luminance values.

As the normal vector distribution has the highest variance $\lambda_1$ along the direction of the first principal vector $v_1$, using the direction of $v_1$ can maximize the overall contrast in the $xy$-plane. In other words, the areas with normal vectors close to either $v_1$ or $-v_1$ will be highlighted, and the areas with normal vectors close to the opposite direction will be shadowed. However, the preferable direction, whether positive or negative, has not been determined and the light direction also needs a proper elevation.
Following O'Shea et al.~\cite{o2008assumed}, we choose the one from the top where the $y$-component is negative and set the elevation to $60^{\circ}$, so the automatic light direction $L_{\textit{auto}}$ is defined as:
\begin{align}
L_{\textit{auto}} = \left\{ \begin{array}{ll}
\text{Combine}(N'_{mean}+\sqrt{\lambda_1}v_1, 60^{\circ}) & \textrm{if $v_1[y] < 0$} \nonumber \\
\text{Combine}(N'_{mean}-\sqrt{\lambda_1}v_1, 60^{\circ}) & \textrm{otherwise}, 
\end{array} \right.
\end{align}
where Combine() generates a 3D unit vector and $v_1$ is scaled, following Zhang et al.~\cite{zhang2013lighting}.
Note that this strategy is also applicable for the exaggerated normal $N_{\textit{DoG}}$ and $L_{\textit{auto}}$ can either be computed over the entire field or the user-selected regions.
\autoref{fig:light_setup:c} shows a shading generated by our automatic lighting setup algorithm (azimuth $=41.44^{\circ}$), where the lighting direction maximizes the overall contrast and clearly shows the ridges and valleys in the blue and green boxes. \autoref{fig:light_setup:d} also reveals the structure, but it does not lead to the most accurate estimations of shapes according to O'Shea et al.~\cite{o2008assumed}.

\cx{This is further supported by comparing the overall variances of shading fields, as the optimal light azimuth should maximize the variations\cite{wambecke2016automatic}. The variance in \autoref{fig:light_setup:c} is 0.0675, which is 18.38\% and 6.15\% higher than the variances in \autoref{fig:light_setup:a} (0.0570) and \autoref{fig:light_setup:d} (0.0636), respectively. Furthermore, the automatic lighting setup outperforms the default configuration in 9 out of 10 tested datasets, with detailed results available in the supplemental material. Additionally, users can interactively adjust the lighting setup if they prefer a different configuration from the automatic one.}

\subsection{Luminance-Only Color Composition}\label{subsect:colorcomposition}
\begin{figure}[!t]
	\centering
	\includegraphics[width=0.99\linewidth]{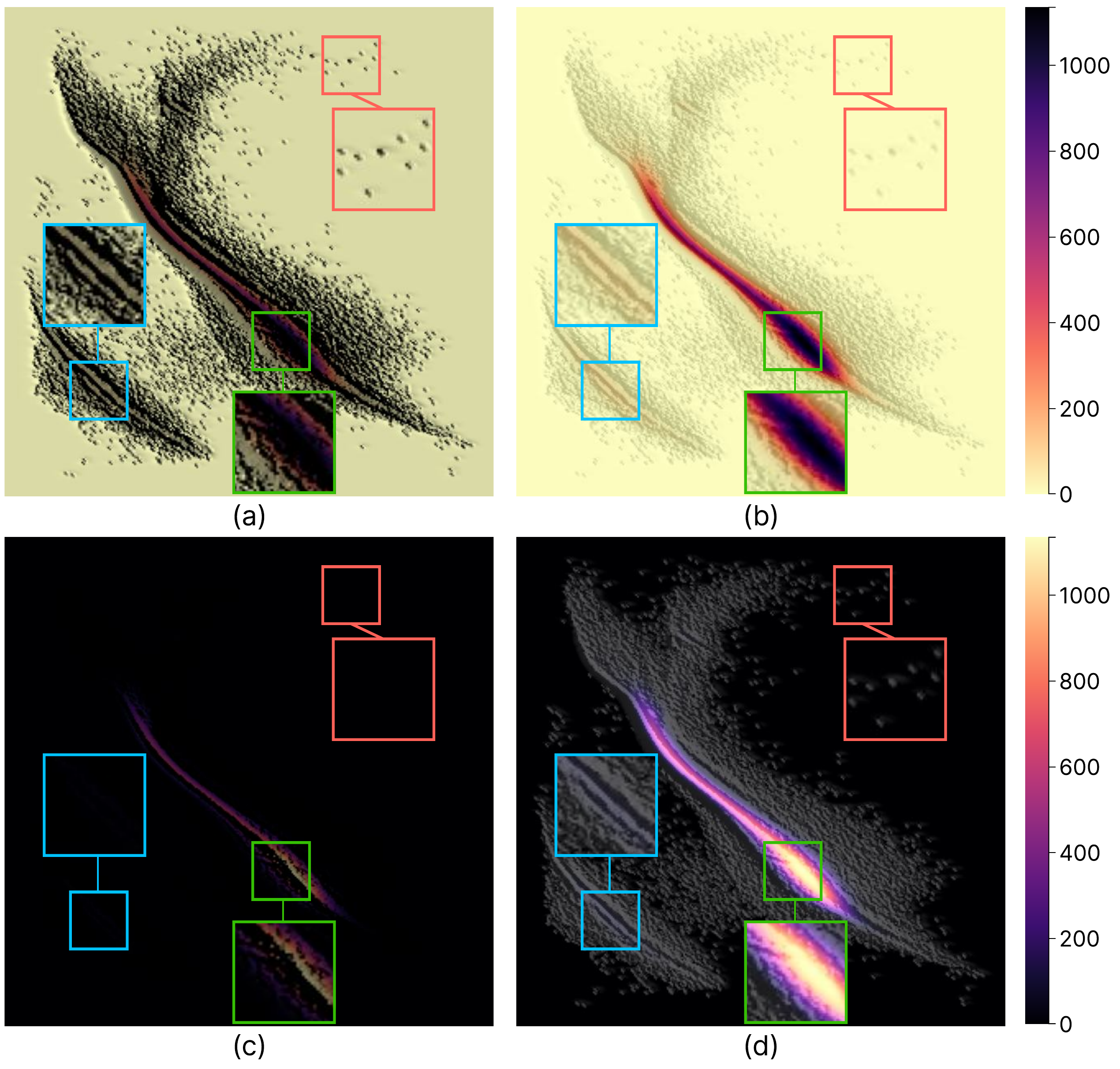}
	\vspace{-6mm}
	\caption{
        Comparison of (a,c) the traditional color composition and (b,d) our luminance-only color composition methods on light and dark backgrounds.
        (a,b) On the light background, the traditional composition reveals local structures but introduces severe color distortions, while our composition preserves the hue and saturation of the original colors when showing structures.
        (c,d) The traditional composition is not feasible on a dark background, while our composition still works well by increasing the luminance according to shading intensities. 
        }
        \begin{subcaptiongroup}
            \phantomcaption\label{fig:color_composition:a}
            \phantomcaption\label{fig:color_composition:b}
            \phantomcaption\label{fig:color_composition:c}
            \phantomcaption\label{fig:color_composition:d}
        \end{subcaptiongroup}
	\label{fig:color_composition}
\end{figure}

After obtaining the shading intensity field $I$, a traditional approach is to naively combine it with 
the density field $F_{\textit{large}}$ through multiplication in the RGB color space:
\begin{align}
I' & = \max(I, 0), \nonumber \\
P & = I' * \text{Colormap}(F_{\textit{large}}),
\label{eq:traditional_composition}
\end{align}
where Colormap() is a one-to-one mapping function from 
density values to RGB values and $P: \mathbb{R}^2 \rightarrow [0,1]^3$ is an image composed of RGB values.
\autoref{fig:color_composition:a} shows the result that combines \autoref{fig:enhancement_pipeline:d} and \autoref{fig:enhancement_pipeline:a} using \autoref{eq:traditional_composition}. Although the structures and outliers in the blue and red boxes are clearly shown, the original colors in the green box and the background are significantly changed, violating \textbf{DR3}.
Furthermore, this composition method is {\em not\/} suitable for dark backgrounds since it always darkens the colors, as shown in \autoref{fig:color_composition:c}.

To address this issue, we introduce a two-step color composition scheme specific to density plot visualizations.
First, we scale the intensity field to fit the luminance range in the CIELAB color space~\cite{schanda2007cie} while ensuring the luminance changes of the empty regions are fixed to 0.
The scaling can be performed using a linear mapping:
\begin{align}
I' & = \phi \cdot \frac{I_{\textit{empty}}-I}{I_{\textit{empty}}-I_{\min}},
\label{eq:intensity_scaling}
\end{align}
where $I_{\min}$ is the global minimum in the field, $I_{\textit{empty}}$ is the intensity of empty regions, \ie $[0,0,1] \cdot L_{\textit{auto}}$, and $\phi$ is a luminance scaling parameter to control the change in luminance corresponding to $I_{\min}$.
The larger the absolute value of $\phi$, the greater the influence of the color composition on the entire plot.
Note that $I_{\textit{empty}}$ never equals to $I_{\min}$ in our scenario, because $L_{\textit{auto}}$ has a fixed elevation of $60^{\circ}$ and always results in a positive $I_{\textit{empty}}$, while $I_{\min}$ is always negative 
since density variations correspond to ridges, which always have shadowed sides under global illumination.
%
Second, we add the scaled intensity field to the luminance channel of the density plot while keeping the other two channels unchanged:
\begin{align}
\text{LAB} & = \text{RGBtoLAB}(\text{Colormap}(F_{\textit{large}})), \nonumber \\
\text{LAB.}L & = \text{clamp}(\text{LAB.}L + I',\ 0,\ 100), \nonumber \\
P & = \text{LABtoRGB}(\text{LAB}). \nonumber
\end{align}
where RGBtoLAB() and LABtoRGB() are conversions between the RGB and CIELAB color spaces, and the clamp() function ensures that any invalid values falling outside the luminance range [0,100] are adjusted to fit within the CIELAB color space.
As shown in \autoref{fig:color_composition:b},\ref{fig:color_composition:d}, our luminance-only composition preserves the appearance of the original colors by maintaining the hue and saturation and works with different backgrounds by flexibly changing the sign of $\phi$. 

Additionally, addition operations are independent of the values of the operands, while multiplication is not. If the original luminance at position $x$ is $l_x$ and the scaled intensity is $i'_x$, the resulting luminance from multiplication will be $l_x i'_x$, which is proportional to the original luminance.
This suggests that the same level of shading intensity will have a more significant impact in areas with higher luminance in the density plot. For color-mapped densities with large luminance, multiplication operations may produce colors not in the colormap of the final visualization (see \autoref{fig:teaser:b} and \autoref{fig:teaser:c}). In contrast, employing additive composition can help reduce color distortions. 

\begin{figure*}[!t]
	\centering
	\includegraphics[width=0.99\linewidth]{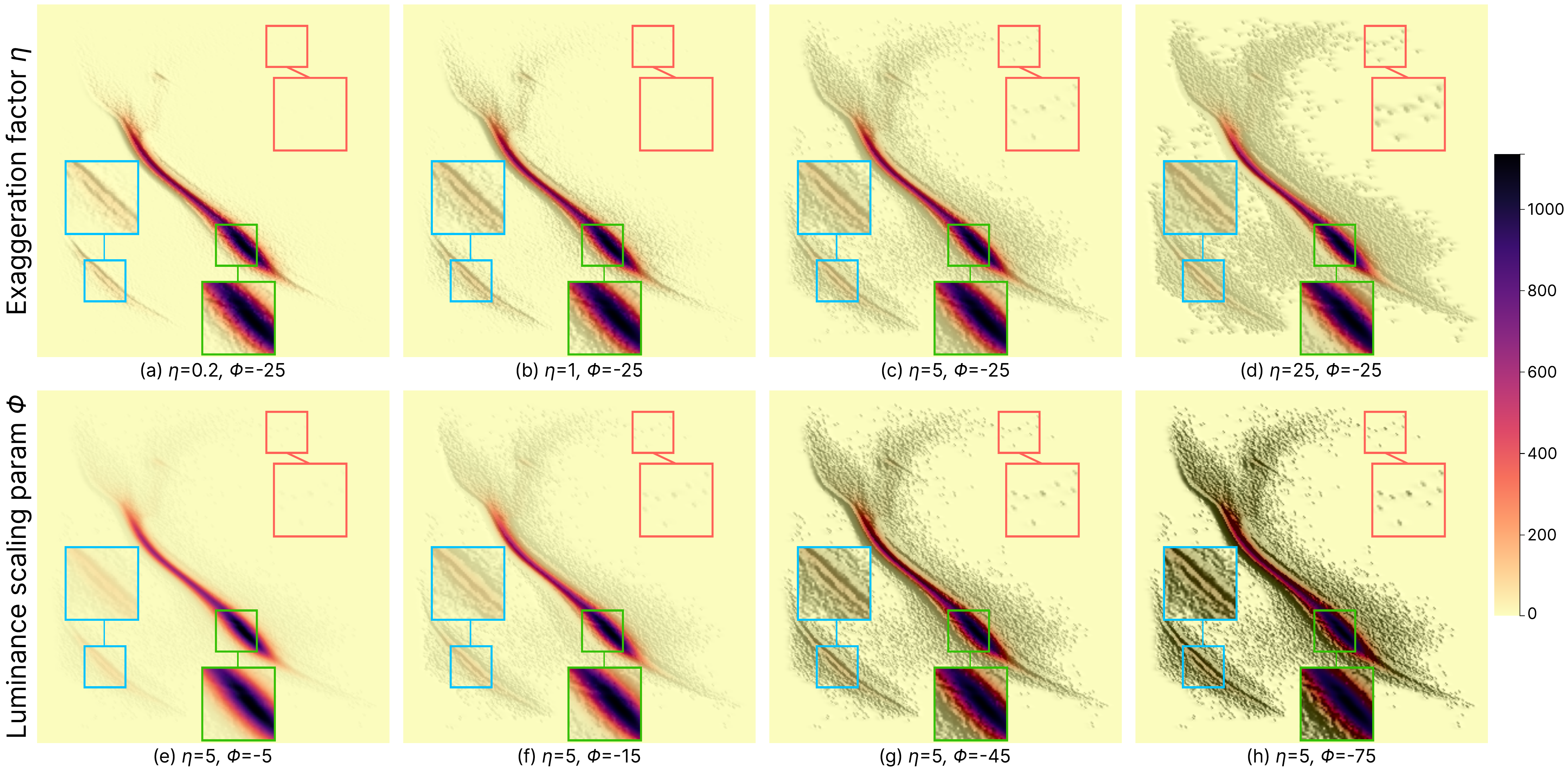}
	\vspace{-2mm}
	\caption{
        Parameter analysis on the \emph{Hertzsprung-Russell diagram} dataset~\cite{babusiaux2018gaia}.
        (a,b,c,d) Increasing $\eta$ makes low-density structures more salient while the high-density structures remain unchanged.
        (e,f,c,g,h) Decreasing $\phi$ makes low-density structures clearer, but at the same time darkens high-density structures.
        }
	\vspace{-7mm}
        \begin{subcaptiongroup}
            \phantomcaption\label{fig:parameter_analysis:a}
            \phantomcaption\label{fig:parameter_analysis:b}
            \phantomcaption\label{fig:parameter_analysis:c}
            \phantomcaption\label{fig:parameter_analysis:d}
            \phantomcaption\label{fig:parameter_analysis:e}
            \phantomcaption\label{fig:parameter_analysis:f}
            \phantomcaption\label{fig:parameter_analysis:g}
            \phantomcaption\label{fig:parameter_analysis:h}
        \end{subcaptiongroup}

	\label{fig:parameter_analysis}
\end{figure*}

\subsection{Parameter Analysis}\label{subsect:parameteranalysis}
In this section, we qualitatively inspect how the parameters of our technique affect the quality of illumination (\autoref{fig:parameter_analysis}). We use a light background and a perceptually uniform colormap, Magma~\cite{hunter2007matplotlib}.
More illumination results under different settings and datasets can be found in the supplementary material.
We also provide a web-based prototype system\footnote{\url{https://xinchen-sdu.github.io/Visualization-Aware-Illumination-for-Density-Plots/}},
in which users can upload datasets
and interactively adjust parameters.

\paragraph{Exaggeration factor $\eta$.}
Parameter $\eta$ determines the degree of exaggeration for the structures in low-density regions.
As the degree of exaggeration increases, the result eventually approaches \emph{aspect shading}, in which the intensity is solely a function of the azimuth of the normal regardless of the density values~\cite{rusinkiewicz2006exaggerated}.
A large $\eta$ leads to clearer outliers, while a small $\eta$ reduces minor variations to reveal major structures.
By default, we set $\eta$ to 5 in order to balance the visibility of outliers and the perception of relative densities based on shading.
The first row of \autoref{fig:parameter_analysis} shows the impact of $\eta$; the small $\eta$ in \autoref{fig:parameter_analysis:a} emphasizes the two bands in the blue box, while the large $\eta$ in \autoref{fig:parameter_analysis:d} improves the visibility of outliers in the red box but makes the depth of structures in the low-density regions look identical compared to \autoref{fig:parameter_analysis:c}.

\paragraph{Luminance scaling parameter $\phi$.}
Parameter $\phi$ influences the amount of modification that can be applied to the luminance channel of the color scale.
For a small value for $\phi$, the luminance changes in the high- and low-density regions are both large, making low-density structures clearer but modifying colors excessively.
In contrast, a large value for $\phi$ maintains the original colors but also weakens the outliers.
However, the background color is retained in all cases to provide the user with a reference to the color map.
We empirically set $\phi$ to -25, which works well for most tested data.
As shown in the second row of \autoref{fig:parameter_analysis}, setting $\phi$ to -5 makes the shadows too faint to show structures, while setting $\phi$ to -75 darkens the colors in the green box and may result in inaccurate perception of the absolute density value.

\section{Evaluation}
We implemented our \ourmethod technique in Python and generated density plots on a PC with an Intel Core i5-4590 3.3GHz CPU and 24GB memory.
To investigate the effectiveness of our approach, we compared it with three density-plot techniques by (i) assessing their color distortions with an established numeric metric~\cite{sharma2005ciede2000} and (ii) conducting an online user study on density-plot-specific analysis tasks. Moreover, two case studies demonstrate the generality and interactivity of our technique.
The raw evaluation materials, including the images of density plots used in the evaluation and the code we used for statistical analysis, can be found in the supplementary material.

\begin{figure}[!t]
	\centering
	\includegraphics[width=0.99\linewidth]{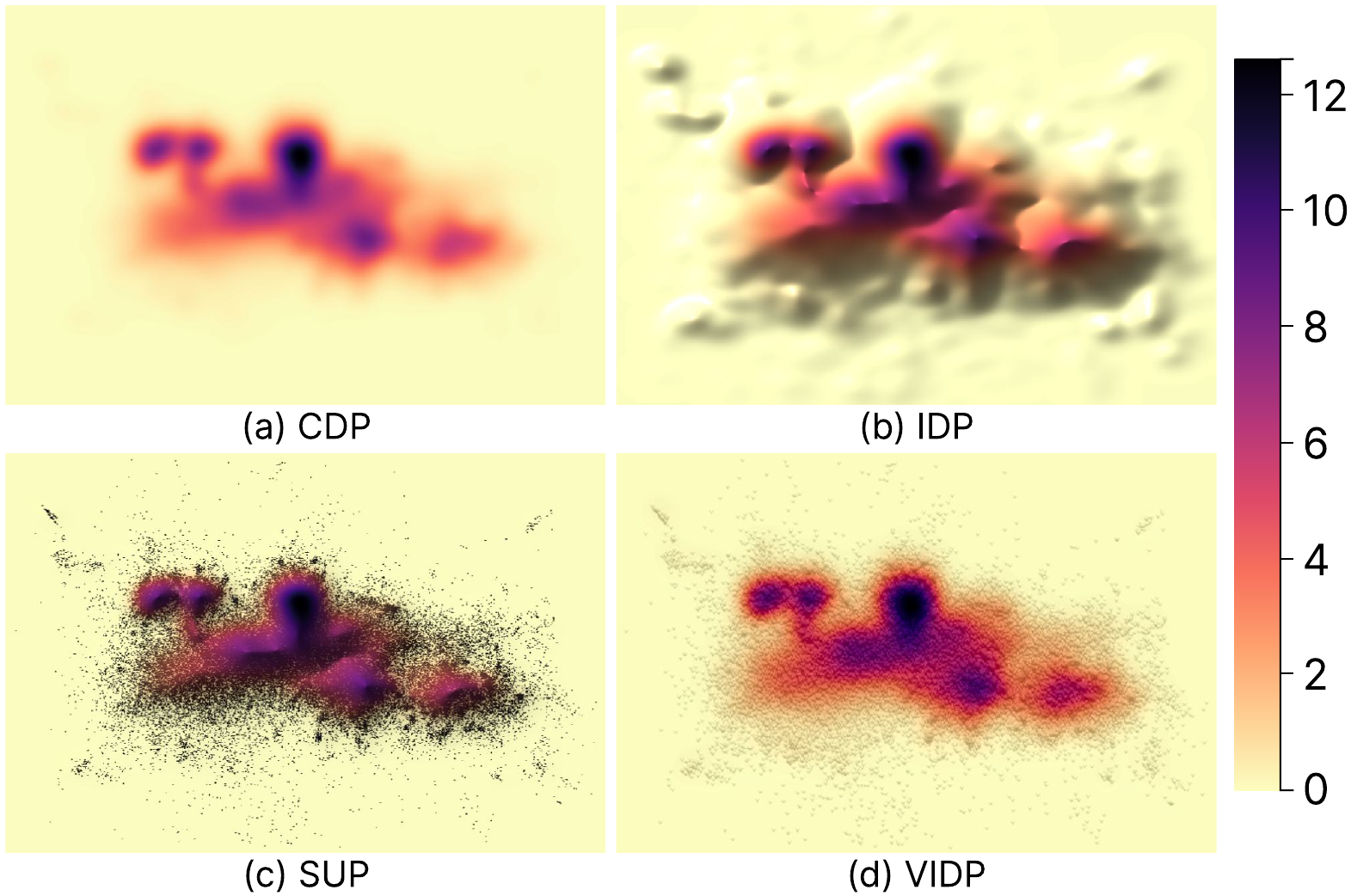}
	\vspace{-5mm}
	\caption{
        Density-plot techniques employed in our user study on the \emph{person activity} dataset~\cite{misc_localization_data_for_person_activity_196}:
        (a) The commonly-used Continuous Density Plot (CDP), (b) Illuminated Density Plot (IDP) using Phong shading, (c) SUnspot Plot (SUP)~\cite{trautner2020sunspot}, and (d) our Visualization-driven Illuminated Density Plot (\ourmethod).
        }
        \begin{subcaptiongroup}
            \phantomcaption\label{fig:density_plot_techniques:a}
            \phantomcaption\label{fig:density_plot_techniques:b}
            \phantomcaption\label{fig:density_plot_techniques:c}
            \phantomcaption\label{fig:density_plot_techniques:d}
        \end{subcaptiongroup}

	\label{fig:density_plot_techniques}
\end{figure}

\vspace{-4mm}
\subsection{Quantitative Evaluation}\label{sec:quantitative_eval}
To confirm that our technique produces less color distortion than the existing ones, we measured the color distortion of the visualization images produced by different techniques.

\paragraph{Density Plot Techniques}.
We compared \ourmethod with three existing density-plot techniques: CDP, IDP~\cite{willems2009visualization}, and SUP~\cite{trautner2020sunspot}, as shown in \autoref{fig:density_plot_techniques}.
Splatterplot (SPP)~\cite{mayorga2013splatterplots} were not included because they were designed not to show density to reduce the visual complexity, and therefore, did not support color-based lookup and comparison tasks.
By default, we set the parameters of our \ourmethod technique to the following values: exaggeration factor $\eta=5$ and luminance scaling parameter $\phi=-25$. The bandwidths of the large and small Gaussian kernels were set by \emph{Silverman’s rule of thumb}, and the reciprocal of the plot width, respectively, and the light direction was automatically chosen as described in~\autoref{sec:auto_light_setup}.
For SUP, we used the default parameters provided by the implementation of the original authors.
For a fair comparison, we set the same bandwidth $h$ for CDP and IDP based on \emph{Silverman’s rule of thumb}, and IDP used the Phong shading model with the same light direction as SUP,~\ie ``top-left'', an azimuth of $120^{\circ}$, and an elevation of $45^{\circ}$.

\paragraph{Datasets}.
For a comprehensive evaluation, we collected ten datasets with substantial diversity in their data distributions and sizes ranging from 4K to 2.5M. 
Half of the datasets were synthetic to ensure the inclusion of outliers. Each synthetic dataset was generated by mixing four Gaussian distributions (for clusters) and a uniform distribution (for outliers). The Gaussian distributions had the same number of points, with randomly-determined means and variances. Outliers were drawn from the uniform distribution covering the range of the four Gaussian clusters, and the number of outliers was set to $0.1\%$ of the points in the clusters.
The other five real-world datasets were collected from the UCI data repository~\cite{blake1998uci} and Kaggle~\cite{kaggle.com} (\autoref{table:real_datasets}). The datasets are available on GitHub\footnote{\url{https://github.com/XinChen-SDU/Visualization-Aware-Illumination-for-Density-Plots}} for replication and as a benchmark for comparing future techniques.
All density plots were generated using the perceptually uniform color map Magma~\cite{hunter2007matplotlib} and were resized to a resolution of $900 \times 600$ pixels.

\paragraph{Measure}.
We utilized CIEDE2000~\cite{sharma2005ciede2000}, a 
standard
perceptual color distance metric, to assess the Degree of Color Distortion (DCD). The images produced by CDP were treated as the baseline because it directly applies the given color map to absolute density values. Color distortion is measured by the average distance against CDP:
\begin{align}
DCD(X,C) = \frac{\sum_{i=1}^N CIEDE2000(X_i,C_i)}{N},
\label{eq:average_ciede2000}
\end{align}
where $X$ is the input image, $C$ is the CDP image on the same dataset, $i$ is a pixel index, and $N$ is the pixel count in $X$.

\begin{figure}[!t]
	\centering
	\includegraphics[width=0.99\linewidth]{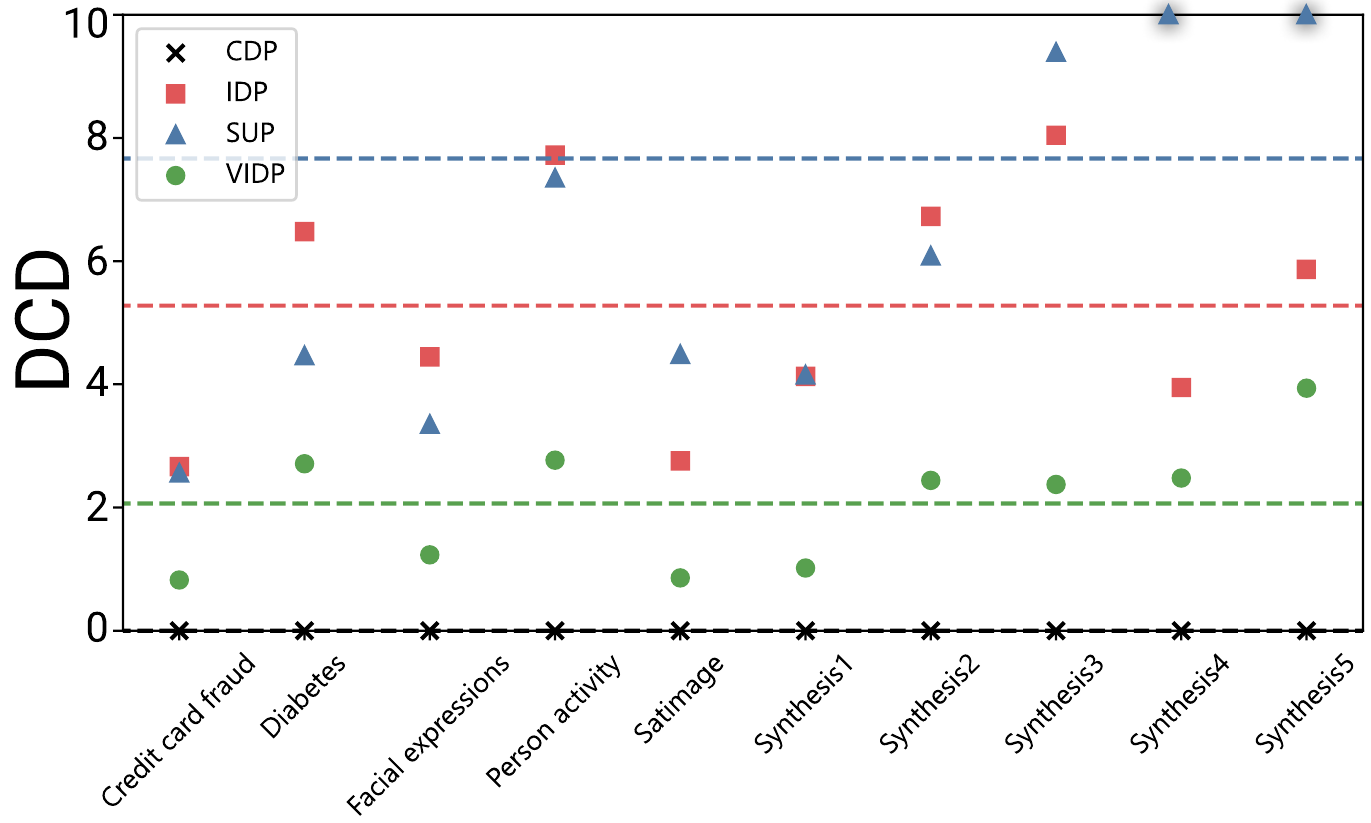}
	\vspace{-6mm}
	\caption{
        Comparison of the degree of color distortion produced by four experimental techniques on ten datasets. Dash lines indicate the average DCD of their corresponding techniques. For values out of the plot range, we treat them as outliers and draw a dark halo to indicate them.
        }

	\label{fig:quantitative_eval}
\end{figure}

\paragraph{Results}.
\autoref{fig:quantitative_eval} shows the results of the quantitative evaluation. With CDP as baseline, \ourmethod has the least color distortion ($\sim$2), followed by IDP ($\sim$5), and SUP performed the worst ($\sim$8).
However, we found IDP performed worse than SUP on a few datasets, such as Person activity~\cite{misc_localization_data_for_person_activity_196}. The reason is that the background color of the medium-density regions of these datasets was largely altered 
by IDP, while SUP showed small scatter points and kept the background color in these regions.
Though \ourmethod modifies the original color to reveal structures, its color distortion is significantly less than IDP and SUP ($p=0.002$ from the Mann-Whitney test against VIDP).

\begin{table}[!t]
	\vspace*{-1mm}
	\centering \resizebox{0.95\linewidth}{!}{\begin{tabular}{@{}l|r@{\hspace{3mm}}r@{\hspace{3mm}}r@{\hspace{3mm}}r@{}}
			\toprule
			Dataset &  \# Points & Max Density & \# Clusters & Outliers (\%)  \\
			\midrule
			{Credit card fraud~\cite{dal2017credit}} & 284,807 & 355.25 & 1 & 0.0007\%  \\
			{Diabetes~\cite{misc_diabetes_130-us_hospitals_for_years_1999-2008_296}} & 99,493 & 13.04 & 2 & 0.0015\%  \\
			{Facial expressions~\cite{misc_grammatical_facial_expressions_317}}  & 12,903 & 2.51 & 7 & 0.0073\%   \\
			{Person activity~\cite{misc_localization_data_for_person_activity_196}} & 98,569 & 12.63 & 7 & 0.0018\%  \\
			{Satimage~\cite{misc_statlog_(landsat_satellite)_146}} & 4,435 & 0.42 & 5 & 0.0577\%  \\
			\bottomrule
			\end{tabular}}
		\vspace*{-1.5mm}
		\caption{
			Summary of the characteristics of the real-world datasets used in our user study. The number of clusters and the proportion of outliers are obtained using the DBSCAN clustering algorithm~\cite{DBSCAN}.
		}
		\label{table:real_datasets}
\end{table}

\begin{figure*}[!t]
	\centering
	\includegraphics[width=\linewidth]{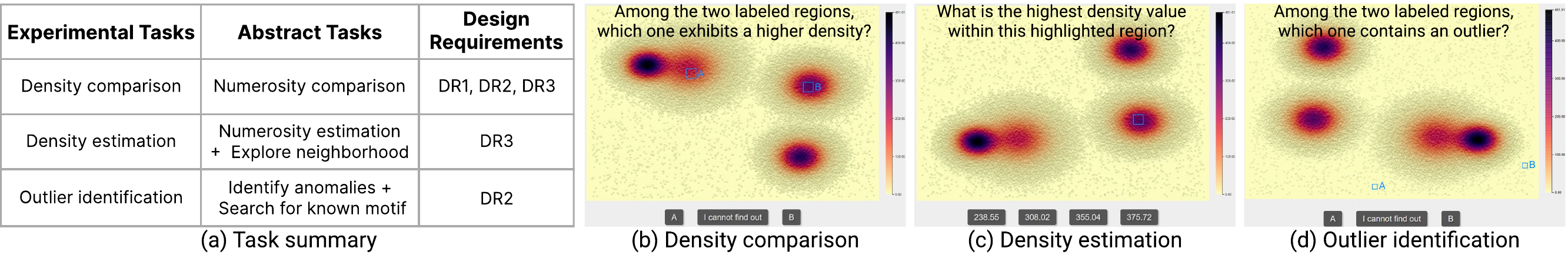}
	\vspace{-6mm}
	\caption{
		The summary of the experimental tasks and exemplified screenshots of the three tasks. (a) The table shows how three experimental tasks correspond to the abstract tasks and design requirements. (b-d) Screenshots and the corresponding questions we asked participants. Images are generated by our \ourmethod technique on a synthesized dataset, where the blue boxes indicating highlighted regions are contained in the stimuli.
		}
	\vspace{-4mm}
          \begin{subcaptiongroup}
            \phantomcaption\label{fig:task_screenshots:a}
            \phantomcaption\label{fig:task_screenshots:b}
            \phantomcaption\label{fig:task_screenshots:c}
            \phantomcaption\label{fig:task_screenshots:d}
        \end{subcaptiongroup}

	\label{fig:task_screenshots}
\end{figure*}

\subsection{Controlled User Study}\label{sec:userstudy}


Our user study compares \ourmethod with three other techniques on three density-plot-specific analysis tasks.

\paragraph{Tasks \& Measures}.
We tested three analytic tasks: \emph{density comparison}, \emph{density estimation}, and \emph{outlier identification}.
These tasks were chosen for three reasons.
First, they are concrete representatives of well-established abstract analytic tasks on scatterplots~\cite{sarikaya2018scatterplots}, covering both aggregate-level tasks and browsing-level tasks, \emph{density comparison} for numerosity comparison, \emph{density estimation} for numerosity estimation and neighborhood exploration, and \emph{outlier identification} for anomaly identification and searching for a known motif (\autoref{fig:task_screenshots:a}).
For this reason, these tasks have also been employed in prior research for scatterplot or density plot evaluation~\cite{yuan2020evaluation, wei2019evaluating, wallner2020multivariate}.
Finally, these tasks are also well aligned with the three design requirements we specified (\autoref{sec:intro}), allowing us to gauge how \ourmethod supports each of the requirements.
A detailed description of the three tasks is given below:

\begin{description}[nosep]
\item[\textbf{T1}:] \emph{Density comparison}: Following the methodology of Trautner et al.~\cite{trautner2020sunspot}, we highlighted two $30\times30$-px regions (A, B) using two blue boxes and asked the participants to choose the region with a higher density (\autoref{fig:task_screenshots:b}). Each question offered three possible answers (A, B, and ``I cannot find out''). 
Only one of A or B was correct, and the last choice was always regarded as a wrong answer.
To measure the error, we scored 0 if the participant’s response was correct and 1 otherwise.

\item[\textbf{T2}:] \emph{Density estimation}: Following the methodology of Trautner et al.~\cite{trautner2020sunspot}, we highlighted one $30\times30$-px region using a blue box and asked the participants to estimate the maximum density value inside the region by looking up the color ramp located on the right side of the scatterplot (\autoref{fig:task_screenshots:c}).
Among four buttons with different density values arranged in ascending order, the participants were asked to click on the button with the correct answer.
The incorrect answers were randomly sampled from a range $[\textit{ans}-\textit{rng}*20\%, \textit{ans}+\textit{rng}*20\%]$, where $\textit{ans}$ is the correct answer and $\textit{rng}$ is the difference between the global maximum and minimum values.
We measured the absolute difference between the participant's answer and the actual maximum density value, divided by the global maximum value to normalize it to [0,1].

\item[\textbf{T3}:] \emph{Outlier identification}: We highlighted two $15\times15$-px sparse regions (A,B) using blue boxes, one of which contains a single data point while the other is empty. The participants were asked to choose the region containing an outlier (\autoref{fig:task_screenshots:d}). Each question offered three possible answers (A, B, and ``I cannot find out'') and only one of A or B was correct. To measure the error, we scored 0 for correct responses and 1 for incorrect.

\end{description}

As shown in \autoref{fig:task_screenshots:a}, the first task is related to \emph{numerosity comparison} and requires the participants to judge relative density, involving both high-, medium-, and low-density regions (\textbf{DR1}, \textbf{DR2}, \textbf{DR3}). 
The second task is about \emph{numerosity estimation} and \emph{explore neighborhood}. Participants have to find the maximum density value within the highlighted region and associate it with the position on the color bar to the right (\textbf{DR3}). The last task involves abstract analysis of \emph{identifying anomalies} and \emph{searching for known motif}, examining the visibility of outliers (\textbf{DR2}).
For all tasks, we did not give an explicit time limit to the participants and measured the response time in seconds.


\vspace{1.5mm}
\paragraph{Stimuli}.
We employed the same techniques and datasets as in~\autoref{sec:quantitative_eval}.
For each dataset and each task, we created two stimuli with different randomly selected non-overlapping highlighted regions. As a result, for each participant, we generated a total of
$3 \textrm{~tasks} \times 4 \textrm{~techniques} \times 10 \textrm{~datasets} \times 2 \textrm{~stimuli} = 240 \textrm{~trials}$.

\vspace{1.5mm}
\paragraph{Hypotheses}.
We expect our approach to outperform the state-of-the-art techniques in preserving relative densities, while simultaneously showing underlying structures and outliers and supporting accurate lookup and comparison of absolute density values. Hence, we postulate the following three hypotheses:
\begin{description}[nosep]
\item[\textbf{H1}:] \ourmethod outperforms the other techniques in terms of accuracy in the \emph{density comparison} task (T1).
\item[\textbf{H2}:] In terms of accuracy, \ourmethod is comparable to CDP in the \emph{density estimation} task (T2), followed by IDP, and SUP has the worst performance.
\item[\textbf{H3}:] In terms of accuracy, \ourmethod is comparable to SUP in the \emph{outlier identification} task (T3), and performs better than CDP and IDP.
\end{description}

\begin{figure*}[!t]
	\centering
	\includegraphics[width=0.98\linewidth]{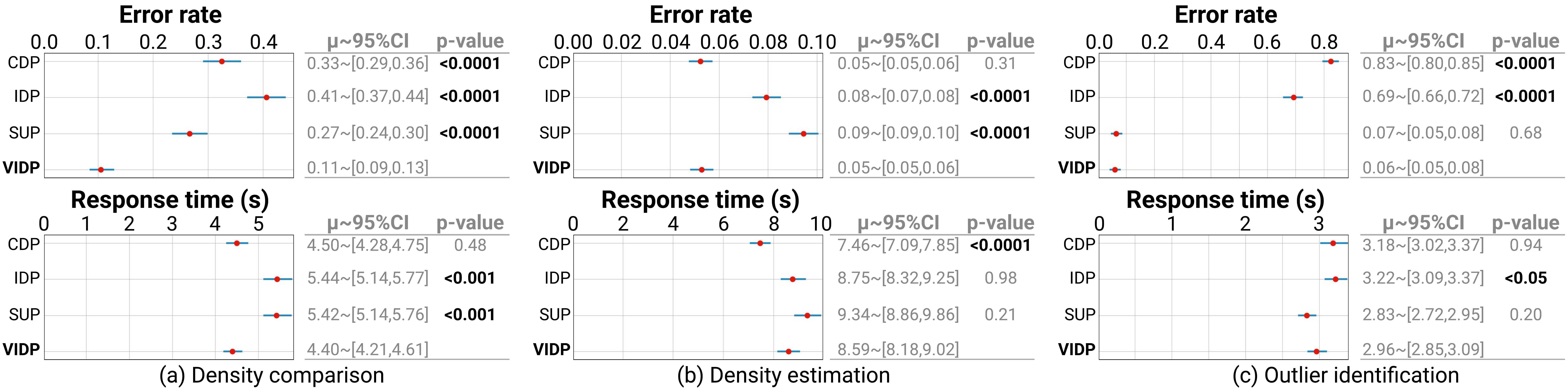}
	\vspace{-2mm}
	\caption{Confidence interval plots and statistical tables of the error rate and response time for each task in our user study.
	Red points indicate the mean values and error bars represent 95\% confidence intervals. Each table shows the statistical test results of our experimental techniques, showing the mean with 95\% confidence interval and p-value from the Mann-Whitney test against our \ourmethod technique.
	}
	\vspace{-6mm}
          \begin{subcaptiongroup}
            \phantomcaption\label{fig:user_study:a}
            \phantomcaption\label{fig:user_study:b}
            \phantomcaption\label{fig:user_study:c}
            \phantomcaption\label{fig:user_study:d}
            \phantomcaption\label{fig:user_study:e}
            \phantomcaption\label{fig:user_study:f}
        \end{subcaptiongroup}

	\label{fig:user_study}
\end{figure*}

\paragraph{Pilot Study}.
We conducted a pilot study with eight graduate students to test our experimental design. Before the study, we explained the tasks to the participants and instructed them to estimate density values based on the color bar to the right.
Then, we only used five real-world datasets, so each participant had to complete $3\ \mathrm{tasks} \times 4\  \mathrm{techniques} \times 5\ \mathrm{datasets} \times 2\ \mathrm{stimuli} = 120$ trials, taking around 15 minutes.
%
We performed a follow-up interview with each participant and asked them if any design factor limited their efficiency and accuracy in completing the task and if they had any suggestions for improving the study.

The pilot study results led us to make three improvements to the initial study design.
First, we identified a learning effect due to the repeated exposure to the five datasets, which allowed some participants to remember the data distributions of specific datasets, e.g., the Person Activity dataset~\cite{misc_localization_data_for_person_activity_196}. To address this issue, we randomly flipped the stimulus vertically or horizontally in each trial while maintaining the highlighted region equivalent. 
Second, for \emph{density comparison} tasks (T1), participants indicated they spent too much time guessing the correct answer, delaying the response time. We added an ``I cannot find out'' button in addition to the two possible answers, allowing them to give up. 
Third, the \emph{density estimation} task (T2) initially asked the participants to enter the exact density value that they estimated, which turned out to be too challenging. After the pilot study, we allow them to choose a correct density value among four choices.
 


\paragraph{Engagement Checks}.
We added a trial for engagement checks every 20 trials in the tasks to ensure participants were engaging with the task.
These trials featured questions that had an obvious answer; for instance, for the \emph{density estimation} task, participants were asked to estimate the density of a nearly-empty region with three among the four choices set to the maximum density value, which were obviously wrong.
We rejected any responses from the participants who failed more than one engagement check. The specific questions used for these trials are detailed in the supplementary material.

\paragraph{Participants}.
We recruited 40 participants from the online research platform \emph{Prolific}, yielding a power of 1 at the effect size of Cohen's f=0.25 and $50\%$ order effect coverage, calculated on Touchstone2~\cite{eiselmayer2019touchstone2}: 23 males and 17 females, aged 18 to 60, including 11 high school students, 19 undergraduates, 8 masters, one Ph.D., and one unspecified. Three participants reported moderate color vision deficiency but passed our engagement checks, so we did not exclude their results.

\paragraph{Procedure}.
We used the Touchstone2 tool~\cite{eiselmayer2019touchstone2} to design a within-subject experiment, in which the order of tasks was fixed (i.e., T1, T2, then T3), and the techniques, datasets, and stimuli were counterbalanced using a Latin square to avoid systematic or random errors~\cite{cox2000theory}.
Each participant went through the following steps on their desktop web browser. 
First, they reviewed a consent page and task instructions.
Second, they watched the tutorial on interpreting the color ramp and completed the three training trials for each task.
Third, complete each trial as accurately as possible, where each participant was asked to take a 2-3 minute break after finishing each task.
Finally, they were asked to provide demographic information. The three training trials were identical to the subsequent real test.
Furthermore, we added four engagement checks for each task to ensure participants were paying attention to the experiment.
On average, the participants took 25 minutes to finish all the trials (min: 12 and max: 42).



\paragraph{Results}.
\autoref{fig:user_study} shows the results for the three tasks.
Following the previous study~\cite{trautner2020sunspot}, we did not assume that the underlying data holds the normality assumption and hence analyzed the results using 95\% confidence intervals using the bootstrap method.
We also performed the Mann-Whitney test for pairwise comparisons between techniques to check if they have significant differences.
Detailed results for each dataset are given in the supplementary material.

\autoref{fig:user_study:a} shows the results of the \emph{density comparison} task (T1). 
Our technique exhibited a significantly lower error rate (0.11) than CDP (0.33), IDP (0.41), and SUP (0.27) ($p<0.0001$).
For response time, \ourmethod (4.40~s) was 
much 
lower than IDP (5.44~s) and SUP (5.42~s) ($p<0.001$) and similar to CDP (4.50~s) ($p=0.48$).
The results confirmed \textbf{H1} that \ourmethod is more accurate than others.


\autoref{fig:user_study:b} shows the results of the \emph{density estimation} task (T2). 
The error rate of \ourmethod (0.05) was comparable to CDP (0.05) ($p=0.31$) and significantly lower than IDP (0.08) and SUP (0.09) ($p<0.0001$).
There were no significant differences in response time between \ourmethod (8.59~s) and existing techniques (IDP: 8.75~s, $p=0.98$; SUP: 9.34~s, $p=0.21$), except CDP (7.46~s) ($p<0.0001$).
The results confirmed \textbf{H2} that \ourmethod does not produce severe color distortions as IDP and SUP do.

\autoref{fig:user_study:c} shows the results of the \emph{outlier identification} task (T3).
There was no statistically significant difference in error rate between \ourmethod (0.06) and SUP (0.07) ($p=0.62$).
Yet, the error rate of our technique was significantly lower than those of CDP (0.83) and IDP (0.69) ($p<0.0001$).
For response time, \ourmethod (2.96~s) was found to be much 
faster than IDP (3.22~s) ($p<0.05$), while the differences with CDP (3.18~s) ($p=0.94$) and SUP (2.83~s) ($p=0.20$) were not statistically significant.
The results confirmed \textbf{H3} that \ourmethod is comparable to SUP and performs better than CDP and IDP for identifying outliers.

In summary, the online study results confirm all our hypotheses and show \ourmethod fulfills the three design requirements, \ie revealing the underlying structures in the given data as much as possible while producing minimal interference with color-encoded density values.

\begin{figure}[!t]
	\centering
	\includegraphics[width=0.98\linewidth]{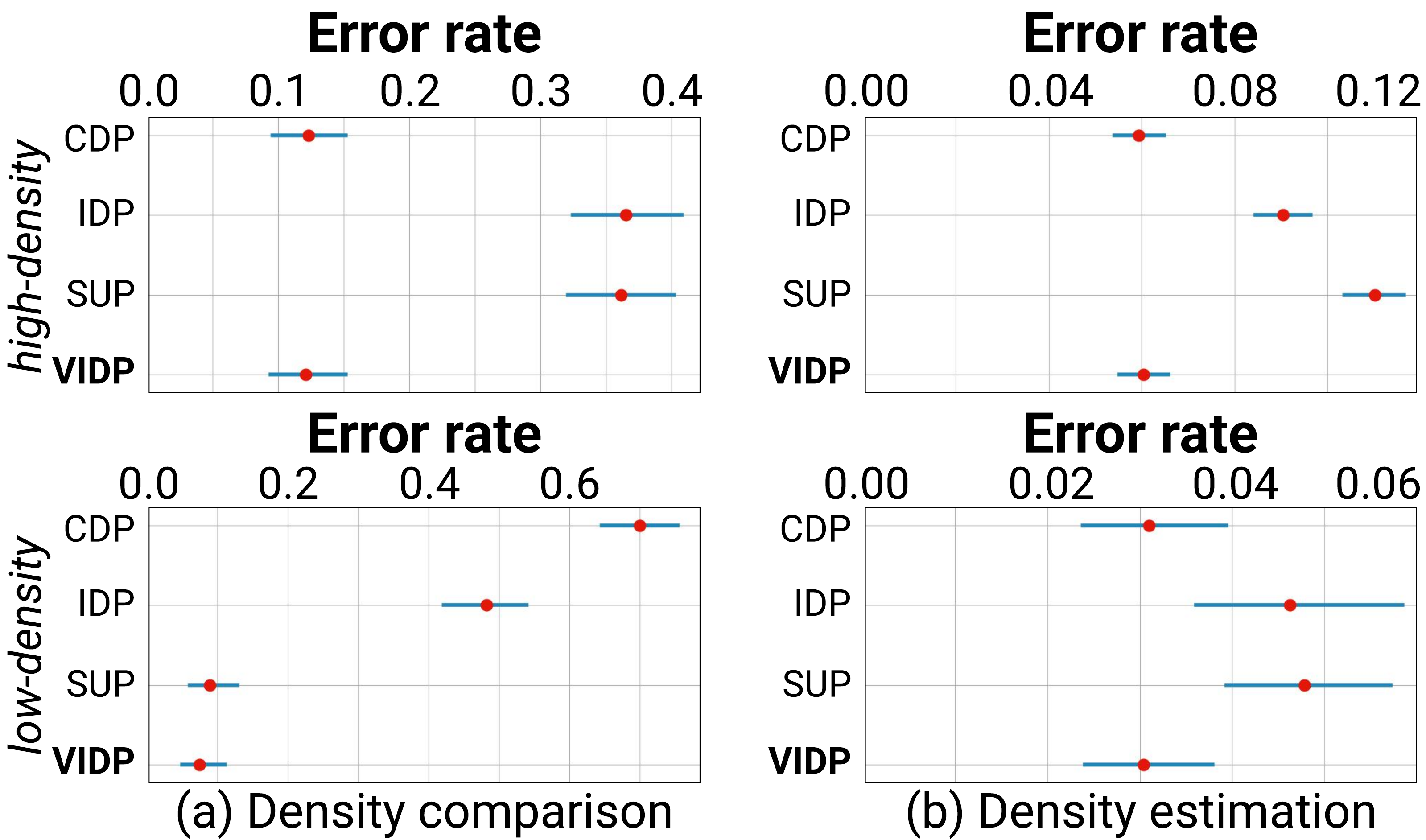}
	\vspace{-1mm}
	\caption{Confidence interval plots of error rates for (a) the \emph{density comparison} and (b) \emph{density estimation} tasks.
	The first and second rows summarize results for stimuli highlighting high- and low-density regions, respectively.
	\ourmethod performs well in all cases.
	}
          \begin{subcaptiongroup}
            \phantomcaption\label{fig:high_low_comparison:a}
            \phantomcaption\label{fig:high_low_comparison:b}
            \phantomcaption\label{fig:high_low_comparison:c}
            \phantomcaption\label{fig:high_low_comparison:d}
        \end{subcaptiongroup}

	\label{fig:high_low_comparison}
\end{figure}

\vspace{1.5mm}
\paragraph{Discussion}.\label{sec:user_study_discussion}
\cx{The above results show that VIDP is able to maintain the original colors and reveal low-density outliers at the same time, and thus provides a more accurate perception of relative data density than existing techniques.}
To further analyze the impact of different techniques on high- and low-density regions, we separate the stimuli in terms of the colors inside the highlighted regions based on CDP. If the difference between any color in the highlighted region(s) and the background color is greater than a JND~\cite{szafir2017modeling}, this stimulus is viewed as high-density, otherwise it is low-density. 
According to this separation, 13 of 20 and 15 of 20 stimuli are high-density in T1 and T2, respectively. In T3, no stimulus is high-density.

\autoref{fig:high_low_comparison} shows the results regarding to the above separation.
As shown in \autoref{fig:high_low_comparison:a}, for the \emph{density comparison} task, \ourmethod performed similarly to CDP ($p=0.89$) and better than IDP and SUP ($p<0.0001$) in high-density stimuli, while performing similarly to SUP ($p=0.41$) and better than CDP and IDP ($p<0.0001$) in low-density stimuli. 
This means that \ourmethod simultaneously supports both color-based comparisons for high-density regions like CDP and structure-based comparisons for low-density regions like SUP.

\autoref{fig:high_low_comparison:b} shows the results for the \emph{density estimation} task. \ourmethod was comparable to CDP and outperformed IDP and SUP in both high-density stimuli (CDP:$p=0.78$, IDP \& SUP: $p<0.0001$) and low-density stimuli (CDP:$p=0.72$, IDP: $p<0.05$, SUP: $p<0.001$).
Since the target of this task is to deduce density values based on color, the results indicate that \ourmethod avoids significant interference with color, while SUP and IDP obscure the original colors and prevent users from accurately perceiving absolute density values.

Because all stimuli are low-density in the \emph{outlier identification} task, \autoref{fig:user_study:c} shows the final results. Although IDP performed slightly better than CDP, participants still made too many errors compared to \ourmethod. This highlights the benefits of the DoGs filter and exaggerated shading employed by our illuminated model for maintaining outliers.

Regarding the response time, our \ourmethod technique is always comparable to or better than IDP and SUP, while CDP only outperforms \ourmethod in the \emph{density estimation} task (T2). By analyzing the participants' feedback, 
we identified two reasons behind the great performance of CDP in T2: (i) the colors of CDP were exactly consistent with the color bar, while all other techniques modified colors to enhance the density plots; and (ii) CDP is a commonly-used technique and the participants were familiar with it before participated in our experiments. 
For the other two tasks, the average response time of \ourmethod is even slightly lower than that of CDP.
Such results confirm that \ourmethod not only enhances density plots on the experimental tasks but also does not impose a high cognitive load, overall.

\begin{figure}[!t]
	\centering
	\includegraphics[width=\linewidth]{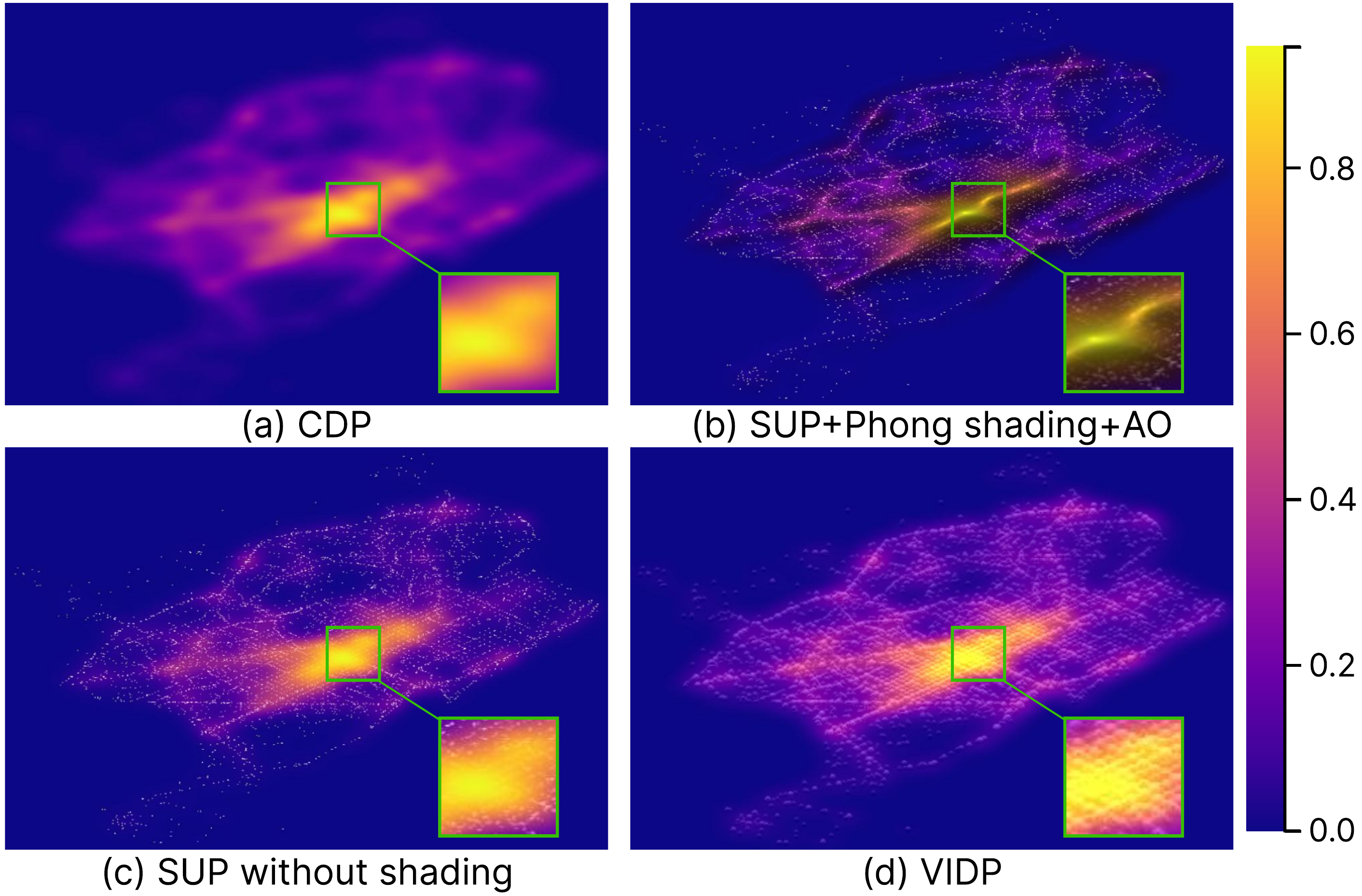}
	\vspace{-4mm}
	\caption{
	 Comparing variants of SUP and our \ourmethod technique using the Barcelona Accidents dataset. (a, b, c) The replicated results of CDP and SUP variants without shading and with Phong shading and ambient occlusion (AO), respectively, as presented in Trautner et al.~\cite{trautner2020sunspot}. (d) The VIDP result with default parameters.  According to~\autoref{eq:average_ciede2000}, the degree of color distortion of (b,c,d) are 4.09, 2.24, and 1.27, respectively.
	}
	 \vspace{0mm}
          \begin{subcaptiongroup}
            \phantomcaption\label{fig:SUP_variants_comparison:a}
            \phantomcaption\label{fig:SUP_variants_comparison:b}
            \phantomcaption\label{fig:SUP_variants_comparison:c}
            \phantomcaption\label{fig:SUP_variants_comparison:d}
        \end{subcaptiongroup}

	\label{fig:SUP_variants_comparison}
\end{figure}

\subsection{Case Studies}\label{sec:Case}
We conducted case studies on two transportation datasets to demonstrate the effectiveness of \ourmethod in a dark background and the local enhancements using our interactive system. An additional case study involving a business dataset is included in the supplemental material.


\paragraph{Barcelona Accidents}.
The results in the previous sections,~\ie Sections~\ref{sec:quantitative_eval} and~\ref{sec:userstudy}, show that SUP performs worse than \ourmethod. Yet, there are two other variants~\cite{trautner2020sunspot}: SUP without shading and SUP with Phong shading and ambient occlusion (AO). 
To learn how these variants perform, we conducted a case study with the 2017 Barcelona Accidents dataset~\cite{kaggle.com}.

\autoref{fig:SUP_variants_comparison:b} demonstrates the same results as those presented in the original paper.
We can see that it reveals structures and outliers, but introduces strong color distortion compared to the CDP result in \autoref{fig:SUP_variants_comparison:a}, especially for the high-density regions. Although the SUP variant without shading in \autoref{fig:SUP_variants_comparison:c} brings less color distortion, it reduces the visibility of structures in high- and medium-density regions (\textbf{DR1}). In contrast, our \ourmethod technique in \autoref{fig:SUP_variants_comparison:d} not only further reduces color distortion but also reveals detailed structures, such as the roads in the green box of~\autoref{fig:SUP_variants_comparison:d}.

\begin{figure}[!t]
	\centering
	\includegraphics[width=0.98\linewidth]{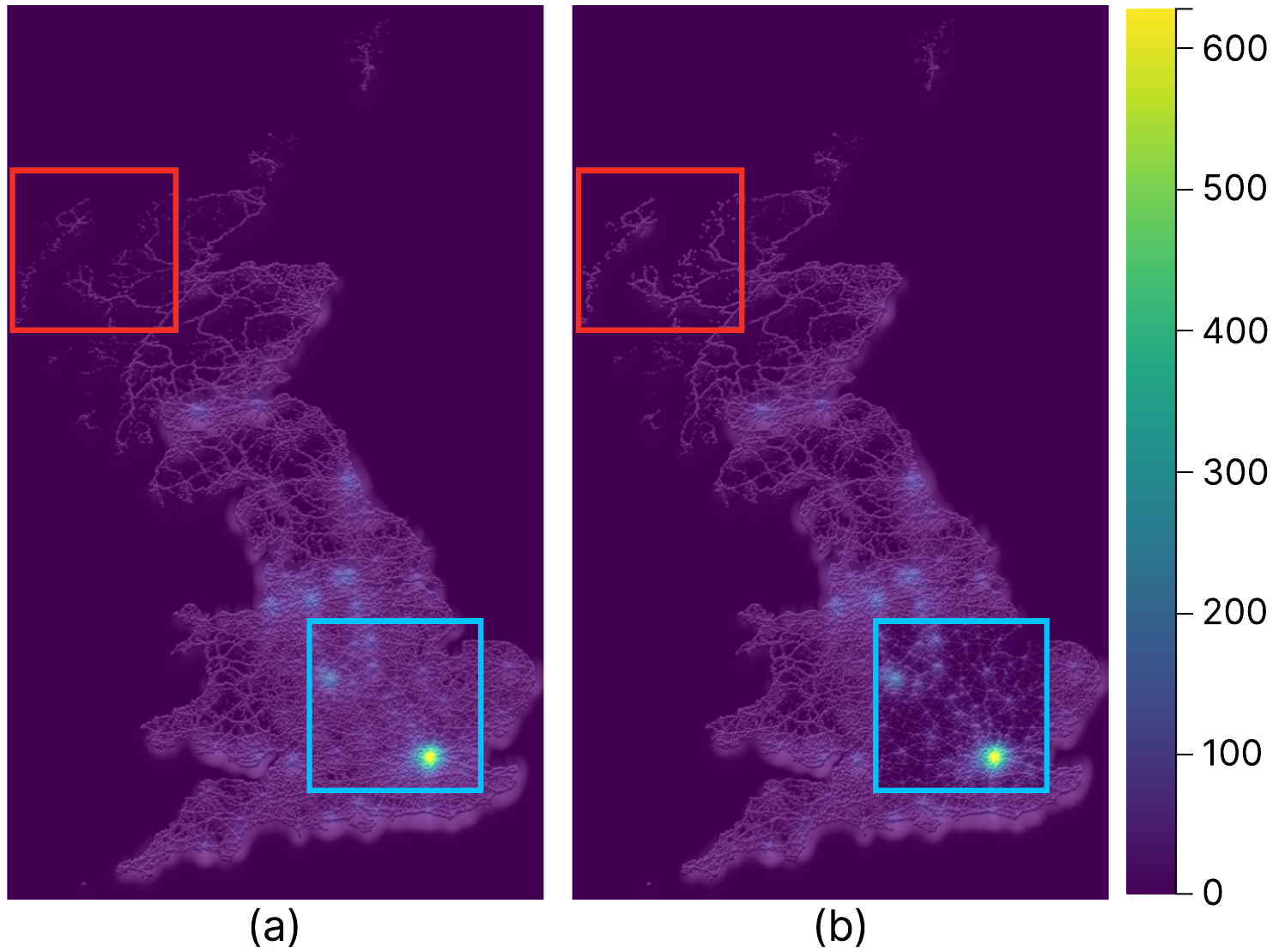}
	\vspace{-2mm}
	\caption{
	\ourmethod of the ``\emph{UK Road Safety}'' dataset~\cite{road_safety} using the Viridis color map and the default parameters. (a) A small local $\eta$=0.2 can suppress minor structures in the area above London (the blue box in (b)), and a large local $\eta$=20 can reinforce minor structures in the northwest area of Scotland (the red box in (b)). 
	}
	 \vspace{2mm}
          \begin{subcaptiongroup}
            \phantomcaption\label{fig:case_study:a}
            \phantomcaption\label{fig:case_study:b}
        \end{subcaptiongroup}

	\label{fig:case_study}
\end{figure}

\paragraph{UK-Road-Safety}.
We used the annual road safety data published by the UK government~\cite{road_safety} from 2005 to 2017 and obtained a density field, containing 2,047,256 records, with the geographical locations (longitude and latitude) mapped to the x and y axes, and the accident frequency at each location mapped to the density.
As shown in \autoref{fig:case_study:a}, we first generated a density plot with default parameters, in which color-encoded densities are preserved and most of the local structures are revealed.
However, the road structures above London (the blue box) and the low-density regions like the northwest area of Scotland (the red box) are not shown clearly.
On the one hand, the road structures above London look blurry because too many minor structures introduce distractions, so we decreased the local $\eta$ to 0.2 to highlight the major roads;
on the other hand, the northwest area of Scotland is low-density, so setting local $\eta$ to 20 helped to reinforce the road structures (\autoref{fig:case_study:b}).

\section{Discussion}
\cx{The evaluations above demonstrate that VIDP outperforms existing density plot designs for \emph{density comparison} and is at least as effective for \emph{density estimation} and \emph{outlier identification}. This advantage stems from our illumination model, which is specifically designed to simultaneously reveal detailed structures and outliers while supporting color-based density estimation. In contrast, CDP is better suited for users focused solely on accurately observing density values, without consideration for local structures or outliers. SUP is effective for those interested in structures and outliers in medium- and low-density regions, provided they are not concerned with the absolute density values represented by colors. IDP performs poorly compared to VDP across all tested tasks, and we recommend against its use except in specific scenarios. Unlike the traditional Phong shading model, our illumination model is visualization-centric, a concept that can also be extended to various chart types, such as line charts~\cite{Wang2023OM3}.}

There are still some limitations in our technique. First, we exploited only the coarse and fine scales of Gaussian fields to compute the structure map.  Doing so limits the ability to convey multi-scale details. We plan to extend our approach to handle multi-scale density fields. Second, although our light setup is data-driven, the parameters $\eta$ and $\phi$ may not be optimal for all datasets. Lastly, we employed simple diffuse shading to present structural information, but our approach is not limited to this traditional model. We will explore the possibility of adopting state-of-the-art learning-based models, such as those described in Jenny et al.~\cite{jenny2020cartographic}, to better characterize meaningful structures for visual analytic tasks. 
\section{Conclusion}
We presented visualization-driven illuminated density plots (\ourmethod), a novel technique to enhance density plots through a visualization-driven illumination model.
We first developed a structure-enhancing shading model, which reveals the detailed structures in high- and medium-density regions and outliers in low-density regions simultaneously to support density-plot-specific analysis tasks.
Then, we introduced a luminance-only color composition method to reduce the color distortion of the original density plot, facilitating more accurate lookup and comparison of absolute density values.
We conducted a quantitative evaluation and a controlled user study to compare our method with the existing density plot techniques, demonstrating that our \ourmethod helps better preserve relative data densities, absolute density values, and outliers.
Also, we presented two case studies to show the robustness of \ourmethod to background colors and the usefulness of interactive local enhancement.

\section*{Acknowledgments}
This work is supported by the grants of the National Key R\&D Program of China under Grant 2022ZD0160805,
NSFC (No.62132017), the Shandong Provincial Natural Science Foundation (No.ZQ2022JQ32), the Fundamental Research Funds for the Central Universities, and the Research Funds of Renmin University of China.




 

\begin{thebibliography}{10}
\providecommand{\url}[1]{#1}
\csname url@samestyle\endcsname
\providecommand{\newblock}{\relax}
\providecommand{\bibinfo}[2]{#2}
\providecommand{\BIBentrySTDinterwordspacing}{\spaceskip=0pt\relax}
\providecommand{\BIBentryALTinterwordstretchfactor}{4}
\providecommand{\BIBentryALTinterwordspacing}{\spaceskip=\fontdimen2\font plus
\BIBentryALTinterwordstretchfactor\fontdimen3\font minus \fontdimen4\font\relax}
\providecommand{\BIBforeignlanguage}[2]{{%
\expandafter\ifx\csname l@#1\endcsname\relax
\typeout{** WARNING: IEEEtran.bst: No hyphenation pattern has been}%
\typeout{** loaded for the language `#1'. Using the pattern for}%
\typeout{** the default language instead.}%
\else
\language=\csname l@#1\endcsname
\fi
#2}}
\providecommand{\BIBdecl}{\relax}
\BIBdecl

\bibitem{sarikaya2018scatterplots}
A.~Sarikaya and M.~Gleicher, ``{Scatterplots: Tasks, Data, and Designs},'' \emph{IEEE Transactions on Visualization and Computer Graphics}, vol.~24, no.~1, pp. 402--412, 2018.

\bibitem{gibson1950perception}
J.~J. Gibson, ``{The Perception of Visual Surfaces},'' \emph{The American journal of psychology}, vol.~63, no.~3, pp. 367--384, 1950.

\bibitem{willems2009visualization}
N.~Willems, H.~Van De~Wetering, and J.~J. Van~Wijk, ``Visualization of vessel movements,'' \emph{Computer Graphics Forum}, vol.~28, no.~3, pp. 959--966, 2009.

\bibitem{trautner2020sunspot}
T.~Trautner, F.~Bolte, S.~Stoppel, and S.~Bruckner, ``{Sunspot Plots: Model-based Structure Enhancement for Dense Scatter Plots},'' \emph{Computer Graphics Forum}, vol.~39, no.~3, pp. 551--563, 2020.

\bibitem{ware2010visual}
C.~Ware, \emph{Visual thinking for design}.\hskip 1em plus 0.5em minus 0.4em\relax Elsevier, 2010.

\bibitem{Phong1975illumination}
\BIBentryALTinterwordspacing
B.~T. Phong, ``{Illumination for Computer Generated Pictures},'' \emph{Commun. ACM}, vol.~18, no.~6, p. 311–317, jun 1975. [Online]. Available: \url{https://doi.org/10.1145/360825.360839}
\BIBentrySTDinterwordspacing

\bibitem{rusinkiewicz2006exaggerated}
S.~Rusinkiewicz, M.~Burns, and D.~DeCarlo, ``{Exaggerated Shading for Depicting Shape and Detail},'' \emph{ACM Transactions on Graphics (TOG)}, vol.~25, no.~3, pp. 1199--1205, 2006.

\bibitem{sharma2005ciede2000}
G.~Sharma, W.~Wu, and E.~N. Dalal, ``{The CIEDE2000 color-difference formula: Implementation notes, supplementary test data, and mathematical observations},'' \emph{Color Research \& Application}, vol.~30, no.~1, pp. 21--30, 2005.

\bibitem{matejka2015dynamic}
J.~Matejka, F.~Anderson, and G.~Fitzmaurice, ``{Dynamic Opacity Optimization for Scatter Plots},'' in \emph{Proceedings of the SIGCHI Conference on Human Factors in Computing Systems}.\hskip 1em plus 0.5em minus 0.4em\relax ACM, 2015, pp. 2707--2710.

\bibitem{micallef2017towards}
L.~Micallef, G.~Palmas, A.~Oulasvirta, and T.~Weinkauf, ``{Towards Perceptual Optimization of the Visual Design of Scatterplots},'' \emph{IEEE Transactions on Visualization and Computer Graphics}, vol.~23, no.~6, pp. 1588--1599, 2017.

\bibitem{silverman1986density}
B.~W. Silverman, \emph{Density Estimation for Statistics and Data Analysis}.\hskip 1em plus 0.5em minus 0.4em\relax London: Chapman \& Hall, 1986.

\bibitem{heer2021fast}
J.~Heer, ``{Fast \& Accurate Gaussian Kernel Density Estimation},'' in \emph{2021 IEEE Visualization Conference (VIS)}.\hskip 1em plus 0.5em minus 0.4em\relax IEEE, 2021, pp. 11--15.

\bibitem{bachthaler2008continuous}
S.~Bachthaler and D.~Weiskopf, ``{Continuous Scatterplots},'' \emph{IEEE Transactions on Visualization and Computer Graphics}, vol.~14, no.~6, pp. 1428--1435, 2008.

\bibitem{mayorga2013splatterplots}
A.~Mayorga and M.~Gleicher, ``{Splatterplots: Overcoming Overdraw in Scatter Plots},'' \emph{IEEE Transactions on Visualization and Computer Graphics}, vol.~19, no.~9, pp. 1526--1538, 2013.

\bibitem{kindlmann2014algebraic}
G.~Kindlmann and C.~Scheidegger, ``{An Algebraic Process for Visualization Design},'' \emph{IEEE Transactions on Visualization and Computer Graphics}, vol.~20, no.~12, pp. 2181--2190, 2014.

\bibitem{10.2312:vmv.20221205}
T.~Trautner, M.~Sbardellati, S.~Stoppel, and S.~Bruckner, ``{Honeycomb Plots: Visual Enhancements for Hexagonal Maps},'' in \emph{Vision, Modeling, and Visualization}, J.~Bender, M.~Botsch, and D.~A. Keim, Eds.\hskip 1em plus 0.5em minus 0.4em\relax The Eurographics Association, 2022.

\bibitem{hughes2014computer}
J.~F. Hughes, \emph{Computer graphics: principles and practice}.\hskip 1em plus 0.5em minus 0.4em\relax Pearson Education, 2014.

\bibitem{gooch2001non}
B.~Gooch and A.~Gooch, \emph{Non-photorealistic rendering}.\hskip 1em plus 0.5em minus 0.4em\relax AK Peters/CRC Press, 2001.

\bibitem{cignoni2005simple}
P.~Cignoni, R.~Scopigno, and M.~Tarini, ``A simple normal enhancement technique for interactive non-photorealistic renderings,'' \emph{Computers \& Graphics}, vol.~29, no.~1, pp. 125--133, 2005.

\bibitem{unwin2006graphics}
A.~Unwin, M.~Theus, and H.~Hofmann, \emph{Graphics of large datasets: visualizing a million}.\hskip 1em plus 0.5em minus 0.4em\relax Springer, 2006.

\bibitem{van2001enridged}
J.~J. van Wijk and A.~Telea, ``{Enridged Contour Maps},'' in \emph{Proceedings Visualization, 2001. VIS'01.}\hskip 1em plus 0.5em minus 0.4em\relax IEEE, 2001, pp. 69--543.

\bibitem{steiger2014visual}
M.~Steiger, J.~Bernard, S.~Mittelst{\"a}dt, H.~L{\"u}cke-Tieke, D.~Keim, T.~May, and J.~Kohlhammer, ``{Visual Analysis of Time-Series Similarities for Anomaly Detection in Sensor Networks},'' in \emph{Computer graphics forum}.\hskip 1em plus 0.5em minus 0.4em\relax Wiley Online Library, 2014, pp. 401--410.

\bibitem{ip2011saliency}
C.~Y. Ip and A.~Varshney, ``{Saliency-Assisted Navigation of Very Large Landscape Images},'' \emph{IEEE Transactions on Visualization and Computer Graphics}, vol.~17, no.~12, pp. 1737--1746, 2011.

\bibitem{lambert1760photometria}
J.~H. Lambert, \emph{Photometria sive de mensura et gradibus luminis, colorum et umbrae}.\hskip 1em plus 0.5em minus 0.4em\relax sumptibus vidvae E. Klett, typis CP Detleffsen, 1760.

\bibitem{horn1981hill}
B.~K. Horn, ``{Hill Shading and the Reflectance Map},'' \emph{Proceedings of the IEEE}, vol.~69, no.~1, pp. 14--47, 1981.

\bibitem{hunter2007matplotlib}
J.~D. Hunter, ``{Matplotlib: A 2D Graphics Environment},'' \emph{Computing in Science \& Engineering}, vol.~9, no.~03, pp. 90--95, 2007.

\bibitem{o2008assumed}
J.~P. O'Shea, M.~S. Banks, and M.~Agrawala, ``{The Assumed Light Direction for Perceiving Shape from Shading},'' in \emph{Proceedings of the 5th Symposium on Applied Perception in Graphics and Visualization}, 2008, pp. 135--142.

\bibitem{zhang2013lighting}
Y.~Zhang and K.-L. Ma, ``{Lighting Design for Globally Illuminated Volume Rendering},'' \emph{IEEE Transactions on Visualization and Computer Graphics}, vol.~19, no.~12, pp. 2946--2955, 2013.

\bibitem{wambecke2016automatic}
J.~Wambecke, R.~Vergne, G.-P. Bonneau, and J.~Thollot, ``{Automatic Lighting Design from Photographic Rules},'' in \emph{WICED: Eurographics Workshop on Intelligent Cinematography and Editing}.\hskip 1em plus 0.5em minus 0.4em\relax Eurographics, 2016, pp. 1--8.

\bibitem{schanda2007cie}
J.~Schanda, ``{CIE} colorimetry,'' \emph{Colorimetry: Understanding the CIE system}, vol.~3, pp. 25--78, 2007.

\bibitem{babusiaux2018gaia}
C.~Babusiaux, F.~van Leeuwen, M.~Barstow, C.~Jordi, A.~Vallenari, D.~Bossini, A.~Bressan, T.~Cantat-Gaudin, M.~Van~Leeuwen, A.~Brown \emph{et~al.}, ``{Gaia Data Release 2 - Observational Hertzsprung-Russell diagrams},'' \emph{Astronomy \& Astrophysics}, vol. 616, p. A10, 2018.

\bibitem{misc_localization_data_for_person_activity_196}
\BIBentryALTinterwordspacing
V.~Vidulin, M.~Lustrek, B.~Kaluza, R.~Piltaver, and J.~Krivec, ``{Localization Data for Person Activity},'' UCI Machine Learning Repository, 2010. [Online]. Available: \url{https://doi.org/10.24432/C57G8X}
\BIBentrySTDinterwordspacing

\bibitem{blake1998uci}
D.~Dua and C.~Graff, ``{UCI Machine Learning Repository},'' \url{https://archive.ics.uci.edu/ml}, 2017.

\bibitem{kaggle.com}
{Kaggle Inc.}, ``Kaggle,'' \url{https://www.kaggle.com/}, 2024.

\bibitem{dal2017credit}
A.~Dal~Pozzolo, G.~Boracchi, O.~Caelen, C.~Alippi, and G.~Bontempi, ``{Credit Card Fraud Detection: A Realistic Modeling and a Novel Learning Strategy},'' \emph{IEEE Transactions on Neural Networks and Learning Systems}, vol.~29, no.~8, pp. 3784--3797, 2017.

\bibitem{misc_diabetes_130-us_hospitals_for_years_1999-2008_296}
\BIBentryALTinterwordspacing
J.~Clore, K.~Cios, J.~DeShazo, and B.~Strack, ``{Diabetes 130-US Hospitals for Years 1999-2008},'' UCI Machine Learning Repository, 2014. [Online]. Available: \url{https://doi.org/10.24432/C5230J}
\BIBentrySTDinterwordspacing

\bibitem{misc_grammatical_facial_expressions_317}
\BIBentryALTinterwordspacing
F.~Freitas, F.~Barbosa, and S.~Peres, ``{Grammatical Facial Expressions},'' UCI Machine Learning Repository, 2014. [Online]. Available: \url{https://doi.org/10.24432/C59S3R}
\BIBentrySTDinterwordspacing

\bibitem{misc_statlog_(landsat_satellite)_146}
\BIBentryALTinterwordspacing
A.~Srinivasan, ``{Statlog (Landsat Satellite)},'' UCI Machine Learning Repository, 1993. [Online]. Available: \url{https://doi.org/10.24432/C55887}
\BIBentrySTDinterwordspacing

\bibitem{DBSCAN}
M.~Ester, H.-P. Kriegel, J.~Sander, and X.~Xu, ``{A Density-Based Algorithm for Discovering Clusters in Large Spatial Databases with Noise},'' in \emph{kdd}, ser. KDD'96.\hskip 1em plus 0.5em minus 0.4em\relax AAAI Press, 1996, p. 226–231.

\bibitem{yuan2020evaluation}
J.~Yuan, S.~Xiang, J.~Xia, L.~Yu, and S.~Liu, ``Evaluation of sampling methods for scatterplots,'' \emph{IEEE Transactions on Visualization and Computer Graphics}, vol.~27, no.~2, pp. 1720--1730, 2020.

\bibitem{wei2019evaluating}
Y.~Wei, H.~Mei, Y.~Zhao, S.~Zhou, B.~Lin, H.~Jiang, and W.~Chen, ``{Evaluating Perceptual Bias During Geometric Scaling of Scatterplots},'' \emph{IEEE Transactions on Visualization and Computer Graphics}, vol.~26, no.~1, pp. 321--331, 2019.

\bibitem{wallner2020multivariate}
G.~Wallner and S.~Kriglstein, ``{Multivariate Visualization of Game Metrics: An Evaluation of Hexbin Maps},'' in \emph{Proceedings of the Annual Symposium on Computer-Human Interaction in Play}, 2020, pp. 572--584.

\bibitem{eiselmayer2019touchstone2}
A.~Eiselmayer, C.~Wacharamanotham, M.~Beaudouin-Lafon, and W.~E. Mackay, ``\emph{Touchstone2}: {An Interactive Environment for Exploring Trade-offs in HCI Experiment Design},'' in \emph{Proceedings of the 2019 CHI Conference on Human Factors in Computing Systems}, 2019, pp. 1--11.

\bibitem{cox2000theory}
D.~R. Cox and N.~Reid, \emph{The theory of the design of experiments}.\hskip 1em plus 0.5em minus 0.4em\relax CRC Press, 2000.

\bibitem{szafir2017modeling}
D.~A. Szafir, ``{Modeling Color Difference for Visualization Design},'' \emph{IEEE Transactions on Visualization and Computer Graphics}, vol.~24, no.~1, pp. 392--401, 2018.

\bibitem{road_safety}
{Department for Transport (DfT), UK}, ``{Road Safety},'' \url{https://www.data.gov.uk/dataset/cb7ae6f0-4be6-4935-9277-47e5ce24a11f/road-safety-data}, 2020.

\bibitem{Wang2023OM3}
\BIBentryALTinterwordspacing
Y.~Wang, Y.~Wang, X.~Chen, Y.~Zhao, F.~Zhang, E.~Wu, C.-W. Fu, and X.~Yu, ``Om3: An ordered multi-level min-max representation for interactive progressive visualization of time series,'' \emph{Proc. ACM Manag. Data}, vol.~1, no.~2, jun 2023. [Online]. Available: \url{https://doi.org/10.1145/3589290}
\BIBentrySTDinterwordspacing

\bibitem{jenny2020cartographic}
B.~Jenny, M.~Heitzler, D.~Singh, M.~Farmakis-Serebryakova, J.~C. Liu, and L.~Hurni, ``{Cartographic Relief Shading with Neural Networks},'' \emph{IEEE Transactions on Visualization and Computer Graphics}, vol.~27, no.~2, pp. 1225--1235, 2020.

\end{thebibliography}
\bibliographystyle{IEEEtran}
\begin{IEEEbiography}[{\includegraphics[width=1in,height=1.25in,clip,keepaspectratio]{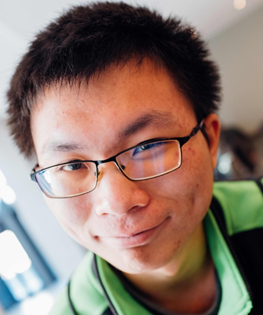}}]{Xin Chen}
is currently a Ph.D. student in the School of Computer Science and Technology at Shandong University.
He obtained his B.E. degree in Software Engineering from Shandong University in 2018.
His research interests include information visualization, human-computer interaction, and data management.
\end{IEEEbiography}

\begin{IEEEbiography}[{\includegraphics[width=1in,height=1.25in,clip,keepaspectratio]{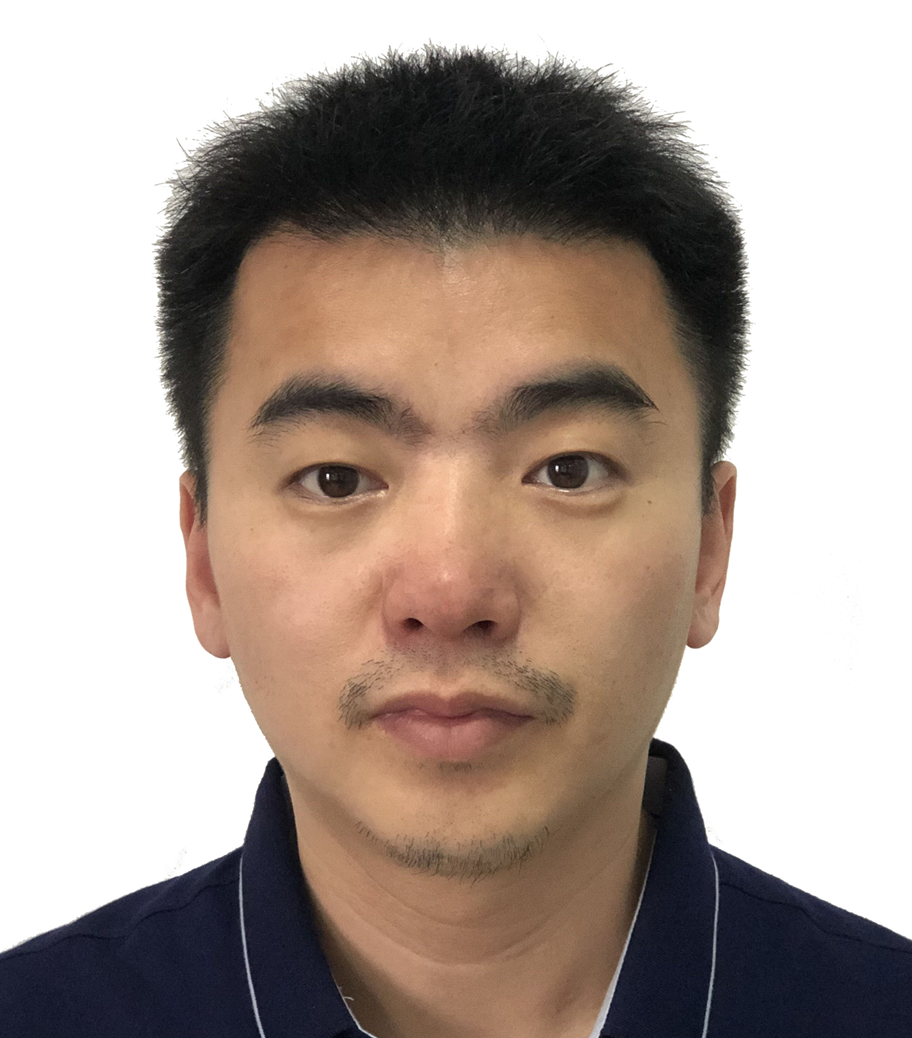}}]{Yunhai Wang}
is a professor in School of Information, Renmin University of China. He serves as the associate editor of IEEE Transactions on Visualization and Computer Graphics, IEEE Computer Graphics and Applications, and Computer Graphics Forum. His interests include scientific visualization, information visualization, and computer graphics.
\end{IEEEbiography}

\begin{IEEEbiography}[{\includegraphics[width=1in,height=1.25in,clip,keepaspectratio]{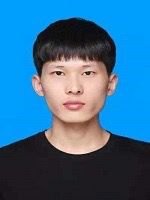}}]{Huaiwei Bao}
is currently a Master student in School of Computer Science and Technology, Shandong University. His research interests include information visualization and database theory.
\end{IEEEbiography}

\begin{IEEEbiography}[{\includegraphics[width=1in,height=1.25in,clip,keepaspectratio]{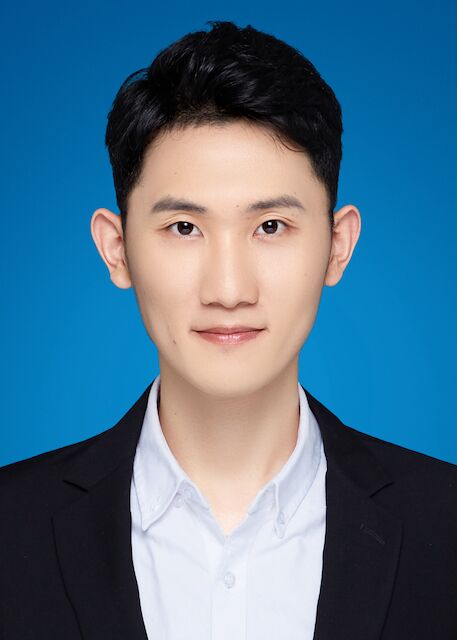}}]{Kecheng Lu}
is a postdoctoral researcher at the School of Information, Renmin University of China. He completed his Ph.D. in Computer Science at Shandong University in 2023. His research focuses on information visualization and human-computer interaction.
\end{IEEEbiography}

\begin{IEEEbiography}[{\includegraphics[width=1in,height=1.25in,clip,keepaspectratio]{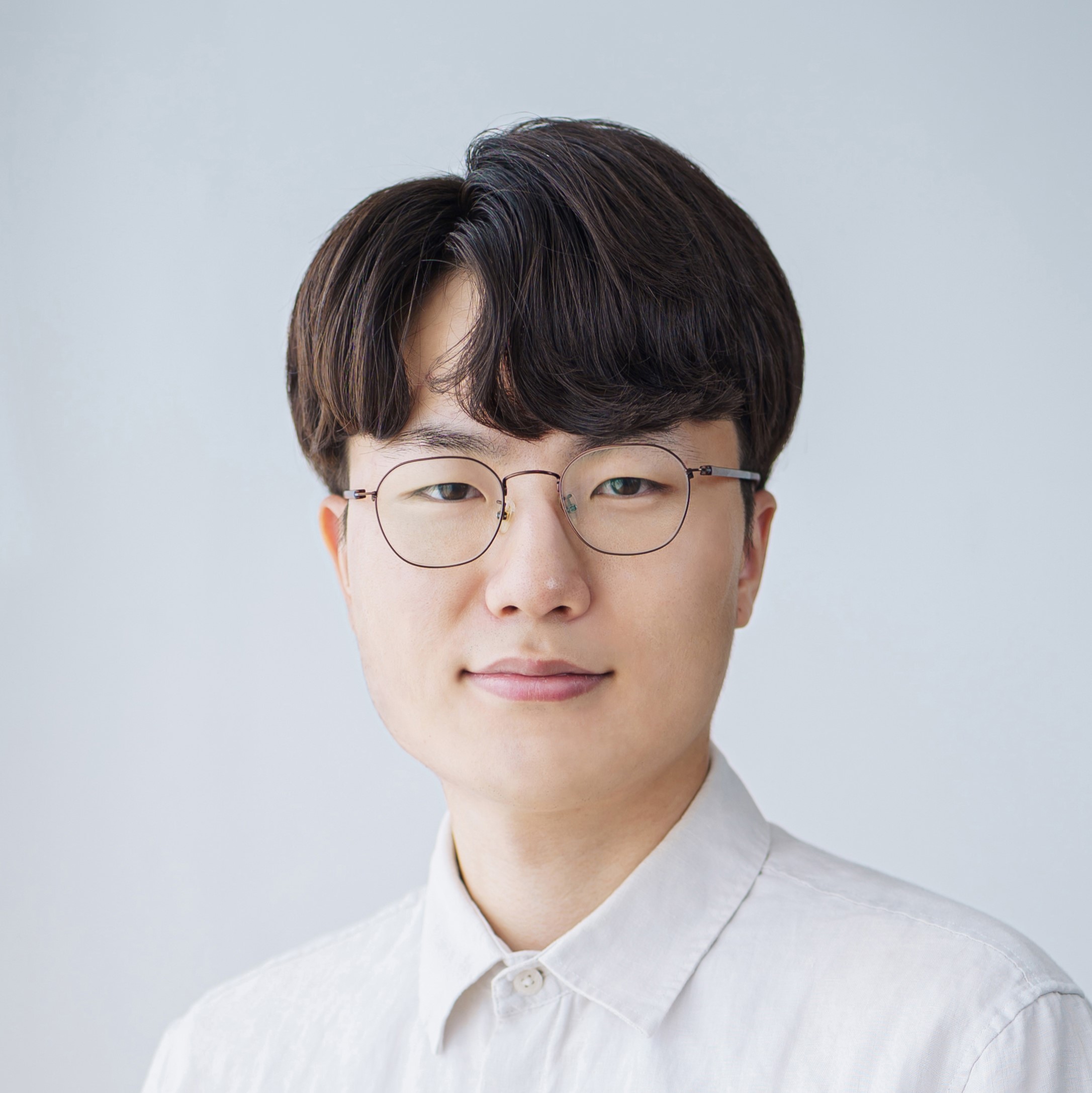}}]{Jaemin Jo}
received the BS and PhD degrees in computer science and engineering from the Seoul National University, Seoul, South Korea, in 2014 and 2020, respectively. He is currently an associate professor with the College of Computing and Informatics, Sungkyunkwan University, Korea. His research interests include human-computer interaction and large-scale data visualization.
\end{IEEEbiography}

\begin{IEEEbiography}[{\includegraphics[width=1in,height=1.25in,clip,keepaspectratio]{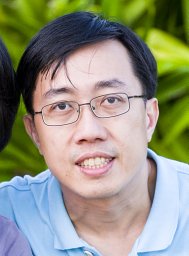}}]{Chi-Wing Fu} is a professor in the Department of Computer Science and Engineering at the Chinese University of Hong Kong (CUHK).  He obtained his PhD from Indiana University, Bloomington.  He served as the program co-chair of SIGGRAPH ASIA 2016 technical brief and poster, associate editor of Computer Graphics Forum, and technical program committee members of SIGGRAPH, IEEE Visualization, etc.  He is currently serving as the Associate Editor-in-Chief of IEEE Computer Graphics and Applications.  His main research interests include computer graphics, 3D vision, user interaction, and visualization.
\end{IEEEbiography}

\begin{IEEEbiography}[{\includegraphics[width=1in,height=1.25in,clip,keepaspectratio]{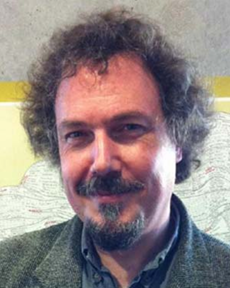}}]{Jean-Daniel Fekete}
(Senior Member, IEEE) received the PhD degree in computer science from the University of Paris-Sud, France, in 1996. He is the scientific leader of the Inria Project Team Aviz that he founded, in 2007. His main research areas are visual analytics, information visualization and human-computer interaction.
\end{IEEEbiography}




\end{document}